\documentclass[11pt]{article}
\usepackage{amsmath,amsfonts,amssymb}
\usepackage[latin3]{inputenc}
\usepackage[dvips]{graphics}
\newtheorem{corollary}{Corollary}
\newtheorem{theorem}{Theorem}
\newtheorem{lemma}{Lemma}
\newtheorem{proposition}{Proposition}
\newtheorem{definition}{Definition}

\newcounter{llista}
\newcounter{llista1}

\begin{document}
\title{Reference frames and rigid motions in relativity}
\author{J Llosa\thanks{e-mail address: pitu@ffn.ub.es} and D Soler\\
       Dept. F\'{\i}sica Fonamental, Universitat de Barcelona,\\ Martí i Franquès,
       1, E-08028  Barcelona, Spain}
\date{15 May 2003}
\maketitle

\begin{abstract}
A reference frame consists of: a reference space, a time scale and a spatial metric. 
The geometric structure induced by these objects in spacetime is developed. 
The existence of a class of spatial metrics that are rigid, have free mobility and can be derived as a slight deformation of the radar metric, is shown. 
\end{abstract}


\section{Introduction \label{S1}}
The notion of reference frame is a fundamental one to any relativity theory. It basically consists of: (i) a reference space, with a given geometry, where a position is assigned to each event and (ii) a time scale to assign a time coordinate to events.

The reference space is based on a reference body, or its ideal extension according to the laws of the appropriate geometry \cite{EinTeo},\cite{EinMean}. By convention, the latter body is rigid: by the very nature of its definition, the reference space is ``what does not change'' and is used as a background to represent the changes in the physical system under study. On its turn, the time scale is implemented by means of a clock or a system of cloks that, by convention, tick at the same rate and are synchronized according to some established protocol.

This way of proceeding does not led to inconsistences, either in Newtonian mechanics or in the special theory of relativity. In a Newtonian reference frame, space is based on an ideal rigid body in an arbitrary state of motion relatively to {\em absolute space}. The time scale is provided by {\em absolute time}. Newton laws of mechanics hold relatively to any frame in this class, provided that the appropriated inertial forces are taken into account.

As for Lorentzian frames in special relativity, the reference space is also one based on a rigid body ${\cal K}$ and its geometry is Euclidean. The time scale is provided by a team of local clocks, identical and synchronized to one another, that are at rest in the reference space. The relative motion of one Lorentzian frame relatively to another ---that is, the motion of the reference body ${\cal K}$ as seen from the reference frame ${\cal K}^\prime$--- is rectilinear and uniform.

If we now go beyond special relativity, either because we are interested in considering arbitrary relative motion or because gravitational interaction is to be taken into account, then things get more complicated, as it is well known from the start of general relativity  \cite{EinMean},\cite{Ein1916},\cite{Ehren}. As illustrated by Ehrenfest paradox, even without a gravitational field, the reference space of a frame ${\cal K}^\prime$ which is in uniform rotational motion relatively to an inertial frame ${\cal K}$ cannot be Euclidean. As a consequence of the equivalence principle, these difficulties in keeping the Euclidean geometry are extensive to the gravitational case. These lines of reasoning lead to resort to a curved geometry for spacetime.

Things turn even worse as far as Born's definition for relativistic ideal rigid bodies \cite{Born1909},\cite{Synge} ---that requires the radar distance between any two infinitely close points keeps constant--- only admits a very restrictive class of relativistic rigid motions, even in Minkowski spacetime \cite{Herglotz}.

The reason of the difficulties commented in the last two paragraphs lies in that, the instantaneous distance between two points $A$ and $B$ in the space of reference, infinitely close to one another, is taken by definition equal to the distance that is measured in a Lorentzian reference frame that sees $A$ and $B$ at instantanous rest.

In order to leave aside the difficulties rised by the above mentioned definition of relativistic rigid motion, without resolving them, it is often argued that it is not an actual problem: the existence of a rigid body would imply the instantaneous transmission of signals, so violating the limit imposed by the speed of light in vacuum. 

However, there are neither real rigid bodies in Newtonian mechanics ---as they would involve infinitely intense forces--- and it does not imply a major trouble. Some real bodies are approximately rigid, and their points approximately follow a {\em rigid motion}, that is, the motion of an ideal rigid body
$$ x^i(t) = s^i(t) + R^i_j(t)\, x_{j0} \,$$
which is the result of composing an arbitrary translational motion $s^i(t)$  and a general rotational motion $R^i_j(t)$. Each rigid motion is determined by the motion of one point whatsoever and the angular velocitiy. 

Then, in order to get a better approximation for the motion of a real solid, the corrections implied by elasticity theory have to be taken into account.

A relativistic extension of the notion of rigid motion would be most useful to approximately describe the kind motion followed by a real body that is ``rigid enough". As it has been commented above, Born's definition does not admit a class of motions wide enough, that is, the motion of one point and its angular velocity cannot be arbitrarily fixed. However, in the dessign of individual experiments the space of reference is taken as ``rigid'', often tacitly. Experimentalists do their best in order to built this space with materials as rigid as possible, whatever that could mean in general relativity. 

Think, for instance, in the experiments were a resonant cavity or an interferometer is used to detect departure from isotropy in the speed of light in vacuum \cite{Brillet}. A Fabry-Perot cavity, made with a material as rigid as possible, is used as an etalon of length. It is besides assumed (also tacitly) that the geometry of the cavity does not change while the experiment lasts. Nor it changes when moved around or rotated. To put it shortly, it is assumed to have the intuitive properties of reference spaces in Newtonian mechanics: it is approximately rigid and approximately fulfills the axiom of free mobility \cite{Cartan}. That is, it approximately represents an ideal notion that has not a proper definition yet in the general theory of relativity.

The aim of the present work is to contribute to a theory of reference frames in general relativity. Our construction is based mainly in the notion of ``reference body''. The worldlines of its points yield a 3-parameter congruence of timelike worldlines filling the region of spacetime that is embraced by the reference frame. We shall denote by $u^\alpha$ the unitary vector field associated to the congruence. Since all events lined along a worldline happen in the same place in the reference space, the notions of timelike congruence and space of reference will be almost used as synonyms.

Another notion that must be precisely stablished is the metric relations between points in the reference space: the distance $d\overline{l}^2$ between the places assigned in space to two infinitely close events, $x^\alpha$ and  $x^\alpha +d x^\alpha$. It is obvious that it must be $d\overline{l}^2 \geq 0$ and that it can be $d\overline{l}^2=0$ only for events happening at the same place, that is on the same worldline in the congruence, or  $d x^\alpha \propto u^\alpha$. The latter will lead us to the notion of {\em spatial metrics} relatively to the congruence ---see definition \ref{d0} below.
A spatial distance between two space points could depend on the instant when it is measured.

In order to implement the main ideas discussed at the begining of the present introduction, we shall besides require that the geometry of space does not change with time and that satisfies the axiom of free mobility. 

Section \ref{S2} is devoted to set the fundamentals of the geometry of timelike congruences in a given spacetime. We first introduce the notion of rigid reference frame (whose reference space is endowed with a Riemannian metric) and see how the metric and the Riemannian connexion in the reference space can be translated into a spatial metric and a linear connexion on the spacetime region swept by the reference space. Later the condition of rigidity for the spatial metric is relaxed and we study a wider class of reference frames, that we label {\em fluid} to emphasize their lack of rigidity. We then see that a linear connexion $\overline{\nabla}$ can also be assigned to the spacetime region embraced by the reference frame. Generally, the latter connexion is not a Riemannian one and is somewhat related with ideas that have been developed somewhere else ---see refs. \cite{Zelm}, \cite{Catt}, \cite{Bel}, \cite{BeLlo} i \cite{Boers}, to cite some few. It will be proved that the connexion $\overline{\nabla}$ is an extension of the rigid case, i. e., if the spatial metric $\overline{g}$ is rigid, the former rigid case is recovered.

Then the notion of subframe of reference is developed, in complete mimicry with submanifolds of a Riemannian manifold \cite{Hicks}. Some results similar to Gauss curvature equation and Codazzi-Mainardi equations ---that will be most usefull to simplify the derivations of section 3--- are obtained. 

We are interested in distinguishing a class of reference frames that could play a role similar to Newtonian reference frames in classical mechanics, wher the relative motion of any two Newtonian frames, ${\cal S}$ and  ${\cal S}^\prime$, either inertial or not, is determined by the motion of the origin $O^\prime$ relatively to ${\cal S}$ and the angular velocity of ${\cal S}^\prime$ relatively to ${\cal S}$.

In our relativistic extension, we would expect that: (a) the congruence of worldlines defining the space of reference is determined when one of its worldlines and the congruence's vorticity on the latter are given, (b) if possible the reference space must be endowed with a Riemannian metric having the free mobility property and (c) in Minkowski spacetime this class of reference frames should include inertial ones.

In spite that it seems to be the natural relativistic extension of the Newtonian rigidity condition, Born's condition \cite{Born1909}, \cite{Synge}, i. e., the preservation of the Fermat metric\footnote{For reasons discussed elsewhere, the name {\em Fermat metric} refers to the metric tensor associated to the radar distance \cite{Bel}} is not a valid rule to characterize the sought class of relativistic reference frames. This is due to the limitations implied by the Herglotz-Noether theorem \cite{Herglotz}.

In an earlier paper \cite{Llo96} one of us proposed a variation of Born's condition, namely, conformal rigidity. It was shown that in a 2+1 dimensional spacetime the class of shear-free motions is wide enough in the sense explained above. Furthermore, a flat 2-dimensional space metric was proved to exist. As an application, the rotational motion of a disk with arbitrary angular velocity in a flat spacetime was unambiguously modelled.  

However the conformal rigidity condition seemed to be too restrictive for a 3+1-dimensional spacetime. We here propose a more sophisticated variation of Born's condition that will yield a satisfactory result in the realistic 3+1 case. 
Instead of requiring that the Fermat metric $\hat{g}$ is preserved (or conformally preserved), we demand that a deformation $\overline{g}$ of the Fermat metric exists having at the same time the properties of rigidity and free mobility. The deformation we propose is what we call a {\em constriction}, that is, a transformation that at every point leaves unchanged the lengths along one space axis and conformally stretches or compresses lengths in the perpendicular plane ---see definitions \ref{d4} and \ref{d4a}.

In section 3 we prove that the class of reference frames that can be associated a spatial metric that: (a) is related to Fermat metric by a constriction transformation, (b) is rigid and (c) has the free mobility property, is large enough. That is, 
for any given timelike worldline and for arbitrarily prescribed values of the vorticvity on this worldline, there is one of such reference frames.

\section{The geometry of timelike congruences \label{S2}}
Let ${\cal V}_4$ be a spacetime domain, $g$ a Lorentzian metric with signature $-+++$ and $u$ a unitary timelike vector field on ${\cal V}_4$:
\begin{equation}
 g(u,u) = -1
\label{e2.1}
\end{equation}
Its integral curves yield a 3-parameter congruence of timelike worldlines ---shortly, a timelike 3-congruence---, ${\cal E}_3$, that allows to establish the following equivalence relation, for $\,x, y \in {\cal V}_4\,,$
$$ x\sim y \Leftrightarrow y \mbox{ lies in the worldline passing through } x \,.$$

The cosets for this equivalence relation are the worldlines belonging to the 3-congruence and:
$$ {\cal E}_3 := {\cal V}_4/\sim $$
is the {\it space of reference} ---shortly, the space of the congruence.

It can be easily proved that it is a 3-manifold. 
The canonical projection $\pi:\,{\cal V}_4 \longrightarrow {\cal E}_3$ is a differentiable map and the tangent and pull-back maps will be respectively denoted by  
$$\pi_\ast:\,T{\cal V}_4 \longrightarrow T{\cal E}_3 \qquad  {\rm and}  \qquad \pi^\ast:\,\Lambda{\cal V}_4 \longrightarrow \Lambda{\cal E}_3\,.$$
It is obvious that:
\begin{equation}
\pi_\ast u = 0 \qquad \mbox{and} \qquad {\rm Ker}\, \pi_\ast = {\rm span}[u] 
\label{e2.2}
\end{equation}

\subsection{The time scale \label{SS2.2}}
A time scale for the congruence ${\cal E}_3$ in the domain ${\cal V}_4$ is a 1-form $\theta \in\Lambda^1{\cal V}_4$ such that: $\langle\theta,u\rangle \neq 0$ everywhere in ${\cal V}_4$. Furthermore, if 
\begin{equation} 
  \langle\theta,u\rangle = -1
\label{e2.3}
\end{equation}
then $\theta$ is called a {\it proper time scale}.

A curve in ${\cal V}_4$ whose tangent vector $v$ satisfies $\langle\theta, v\rangle =0$ connects events that are simultaneous according to the time scale $\theta$ ---shortly, $\theta$-simultaneous.

In particular, combining the Lorentzian metric $g$ and the unitary vector field $u$ that defines the congruence, we obtain the {\it Einstein proper time scale} $\theta_0:=g(u,\_ )$, that is,
\begin{equation}
\langle\theta_0,v\rangle := g(u,v) \,, \qquad \forall v\in{\cal V}_4
\label{e2.4}
\end{equation}
This time scale is physically based on a system of identical local clocks, ticking local proper time and synchronized according to Einstein's protocol \cite{Ein1905} .

Once a time scale $\theta$ is chosen, any vector $v\in T{\cal V}_4$ can be unambiguously split as:
\begin{equation}
v = V - \langle\theta, v\rangle u \qquad \mbox{with} \qquad V\in {\rm Ker}\,\theta 
\label{e2.5}
\end{equation}
that is, a component $V$ which is orthogonal to $\theta$ and a time component (parallel to $u$). Thus, at any $x\in {\cal V}_4$ we have the canonical splitting:
\begin{equation}
   T_x{\cal V}_4 = {\rm Ker}\,\theta_x \oplus {\rm span}[u_x] 
\label{e2.6}
\end{equation}
The vectors in ${\rm Ker}\,\theta_x$ are called {\it spatial}\footnote{Relativetly to the given congruence and time scale}.

An analogous splitting also applies to 1-forms $\alpha\in T^\ast{\cal V}_4$:
\begin{equation}
\alpha = \alpha_\perp - \langle\alpha, u\rangle \theta \qquad \mbox{with} \qquad \langle\alpha_\perp,u\rangle=0 
\label{e2.7}
\end{equation}
which corresponds to
$$   T^\ast_x{\cal V}_4 = [u_x]^\perp \oplus {\rm span}[\theta_x] $$
1-forms in $[u_x]^\perp$ are called {\it spatial}. 

It is interesting to notice that the spatial components of a given 1-form $\alpha$ and a given vector $v$ meet the following relation:
$$ \langle\alpha_\perp,v \rangle = \langle\alpha,V\rangle  $$

The above splitting of vectors and 1-forms also extends to tensors of any order. Consider for instance $ M$, a 1-covariant, 1-contravariant tensor on $T{\cal V}_4$. We can separate it as:
\begin{equation}
M = M_\perp - u\otimes \beta_\perp - W\otimes \theta + m_0 u \otimes\theta 
\label{e2.7a}
\end{equation}
where $M_\perp$, $\beta $ and $W$ are all spatial and, for any given $\alpha$ and $v$, 
\begin{equation}
M_\perp(\alpha,v) := M(\alpha_\perp, V) \;,\qquad  \langle\beta,v\rangle := M(\theta,V) \;,\qquad \langle\alpha,W\rangle := M(\alpha_\perp,u)   \;,\qquad  m_0:= M(\theta,u)  
\label{e2.7b}
\end{equation}

In terms of a basis of vectors $\{e_1,e_2,e_3,e_4\}$ for $T_x{\cal V}_4$ and its dual basis $\{\omega^1,\omega^2,\omega^3,\omega^4\}$ for $T^\ast_x{\cal V}_4$, we have: 
$$ u = u^\alpha e_\alpha  \qquad \qquad  \theta = \theta_\alpha \omega^\alpha $$
Then the components of $V$  and $\alpha_\perp$ in equations (\ref{e2.5}) and (\ref{e2.7}) respectively are
$$ V^\mu = P^\mu_\rho \, v^\rho \qquad {\rm and} \qquad (\alpha_\perp)_\mu = P^\mu_\rho \,\alpha_\mu \,,$$
where
\begin{equation}
P^\mu_\rho := \delta^\mu_\rho + u^\mu\theta_\rho
\label{e2.8}
\end{equation}
is the projector orthogonal to $u$ and $\theta$ or {\em spatial projector}. 

\begin{definition} \label{d1}
A vector field $v$ on ${\cal V}_4$ is said (spatially) {\it projectable} when, for $x, y\in {\cal V}_4$,
$$ \pi x = \pi y \quad \Rightarrow \quad \pi_\ast v_x = \pi_\ast v_y $$
\end{definition}

\begin{proposition} \label{p1}
A vector field $v$ is projectable if, and only if, $\pi_\ast[u,v] =0$
\end{proposition}
where $[u,v]={\cal L}(u)$ is the Lie derivative.

The proof is based in that, if  $\forall p=\pi x\in {\cal E}_3$, then $\forall f\in\Lambda^0{\cal E}_3$ we have that:
$$ \left(\pi_\ast[u,v]\right)_p f = [u,v]_x(f\circ\pi) = u_x\left((\pi_\ast v)f\right) $$
where (\ref{e2.2}) has been used. Now, let $\varphi_tx$ be the worldline in the congruence passing through $x$, with $\varphi_0x=x$, then
$$ \left(\pi_\ast[u,v]\right)_p f =u_x\left((\pi_\ast v)f\right)= \frac{d\;}{dt} \left\{\pi_\ast\left(v_{\varphi_tx}\right)f\right\}_{t=0} $$
If $v$ is projectable, then $\displaystyle{\pi_\ast\left(v_{\varphi_tx}\right)}$ does not depend on $t$, and therefore $\pi_\ast[u,v] =0$.

Conversely, if $\pi_\ast[u,v] =0$, then $\displaystyle{\pi_\ast\left(v_{\varphi_tx}\right)}$ is constant on the worldline passing through $x$, which means that, according to definition \ref{d1}, $v$ is projectable. \hfill $\Box$

\begin{proposition} \label{p2}
Let $\vec{v}$ a vector field tangent to ${\cal E}_3$, then there exists a unique vector field $V$ tangent to ${\cal V}_4$ which is projectable and spatial, that is
$$\pi_\ast V= \vec{v} \qquad \mbox{and }\qquad \langle\theta,V\rangle =0 \,.$$ 
\end{proposition}

\paragraph{Proof: }
Consider an open covering $\{U_i\}$ of ${\cal E}_3$, such that $\pi^{-1}U_i\subset{\cal V}_4$ admits a submanifold ${\cal S}_i$ that intersects each worldline in the congruence at only one point. Now, applying Proposition III.2.12 in ref. \cite{Godbi}, there exists a finer partition of unity of ${\cal E}_3$, $\{V_j,a_j\}$. Hence $\pi^{-1}V_j$ admits a section ${\cal S}^\prime_j$ that intersects any worldline at one point at most. 

When restricted to ${\cal S}^\prime_j \subset \pi^{-1}V_j$, the projection map $\pi$ is 1 to 1. Hence, there exists a vector field $v_{j0}$ defined at every point $p\in {\cal S}^\prime_j$ by $\displaystyle{v_{j0\,p} = \pi_\ast^{-1}(\vec{v}_{\pi p})}$. This vector field $v_{j0}$ can then be extended to the whole $\pi^{-1}V_j$ by the transport law: ${\cal L}(u)v_j=0$.

Now, $\{\pi^{-1}V_j,a_j\circ\pi\}$ is also a partition of unity of ${\cal V}_4$. We then define
$v = \sum (a_j\circ\pi)\, v_j$. It obviously follows that 
$$\pi_\ast v = \sum (a_j\circ\pi)\, \vec{v} = \vec{v} \,.$$

The sought spatial, projectable vector field is then: $V = v + \langle \theta,v\rangle \,u$ 

To prove the uniqueness, it suffices to notice that $\pi_\ast V_1 =\pi_\ast V_1$ implies that $V_1-V_2 \in {\rm span}[u]$. If we further use that $\langle\theta, V_1\rangle = \langle\theta, V_2\rangle$, we conclude $V_1=V_2$. \hfill $\Box$

\subsection{Rigid reference frames\label{SS2.3}}
A {\it rigid reference frame} is defined by the triple $({\cal E}_3,\theta,g_3)$ ---a timelike 3-congruence, a proper time scale and a Riemannian metric on the reference  space. $\stackrel{3}\nabla$ will denote the Riemannian connexion on $({\cal E}_3,g_3)$.

The Riemannian metric $g_3$ on the reference space ${\cal E}_3$, can then be pulled back to a metric $\overline{g}:=\pi^\ast g_3$ on ${\cal V}_4$, so that the $\overline{g}$-metric product between any couple of vector fields tangent to ${\cal V}_4$
\begin{equation}
\overline{g}(v,w) = g_3(\pi_\ast v,\pi_\ast w)\circ \pi
\label{e2.9}
\end{equation}
is a smooth function.

The metric $\overline{g}$ is degenerated and 
\begin{equation}
{\rm Rad}(\overline{g}) = {\rm span}[u]
\label{e2.10}
\end{equation}
as it follows from (\ref{e2.2}) and (\ref{e2.9}). Moreover, the $\overline{g}$-product of any two vector fields $v$ and $w$ on ${\cal V}_4$ is non-negative:
\begin{equation}
\overline{g}(v,w) \geq 0
\label{e2.10a}
\end{equation}

\begin{proposition} \label{p3}
The metric $\overline{g}$ is rigid: $\,{\cal L}(u) \overline{g} =0$.
\end{proposition}

\paragraph{Proof: }
Given $v_p,\,w_p\in T_p{\cal V}_4$, let $v$ and $w$ be two vector fields tangent to ${\cal V}_4$ such that: they take these given values at $p$ and $[u,v]=[u,w]=0$. Then, using (\ref{e2.9}) and (\ref{e2.2}), we have that:
\begin{eqnarray*}
\hspace*{2em}
\left.{\cal L}(u)\overline{g}(v,w)\right|_p &=& u\left(\overline{g}(v,w)\right)_p - \overline{g}\left([u,v],w\right)_p-\overline{g}\left(v,[u,w]\right)_p \\
  & = & u\left\{g_3(\pi_\ast v,\pi_\ast w)\circ\pi\right\}_p = \pi_\ast u\left\{ g_3(\pi_\ast v,\pi_\ast w)\right\}_{\pi p} = 0 \hspace*{6em} \Box
\end{eqnarray*}

In what follows, it will be interesting to consider a special class of spacetime metrics, having the properties (\ref{e2.10}) and (\ref{e2.10a}), no matter whether they are rigid or not.

\begin{definition} \label{d0}
A symmetric 2-covariant tensor field $\overline{g}$ on ${\cal V}_4$ is said to be {\it a spatial metric} relatively to the congruence ${\cal E}_3$ if:
\begin{list}
{(\alph{llista})}{\usecounter{llista}}
\item{ \hspace*{1em}${\rm Rad\,}(\overline{g}) = {\rm span}[u]$ \, and  }
\item{\;for all $v$, $w$, vector fields on ${\cal V}_4$, $\overline{g}(v,w) \geq 0\,$, and it is  $\overline{g}(v,w) \geq 0\,$ only if one of the vectors is proportional to $u$.}
\end{list}
\end{definition}

The Fermat metric is an instance of this kind of spatial metrics:
$$\hat{g} := g + \theta_0\otimes\theta_0 \qquad{\rm with} \qquad \theta_0:=g(u,\_ )$$
whose expression in a basis is: $\hat{g}_{\alpha\beta } =g_{\alpha\beta }+u_{\alpha}u_{\beta }$. In general, it is not rigid \cite{Herglotz}.

\begin{lemma}  \label{l1}
Let $\overline{g}$ be a spatial metric relatively to ${\cal E}_3$, then for any $x\in {\cal V}_4$, $\overline{g}_x$ is non-degenerate and positive when restricted to ${\rm Ker}\, \theta_x$, that is, to spatial vectors in $T_x{\cal V}_4$.
\end{lemma}

Indeed, let $V_x \in {\rm Ker}\, \theta_x$ be such that $\forall W_x \in {\rm Ker}\, \theta_x$\,, $\; \overline{g}_x(V_x,W_x) = 0$. From (\ref{e2.6}) and (\ref{e2.10}) it follows that $V_x \in {\rm Ker}\, \theta_x \cap {\rm span}[u_x]$, which is $\{0\}$ because $\langle\theta , u\rangle \neq 0$. Hence, $\overline{g}_x$ is non-degenerate in ${\rm Ker}\, \theta_x$. The positiveness follows immediately form (\ref{e2.10a}). \hfill $\Box$

The following result is the converse of Proposition \ref{p3}:

\begin{theorem} \label{t1}
Let $\overline{g}$ be a spatial metric relatively to ${\cal E}_3$, such that ${\cal L}(u)\overline{g} =0$, then it can be unambiguously projected onto a Riemannian metric $g_3$ on ${\cal E}_3$, i. e.,
$$ \exists \mbox{ a metric}\quad g_3 \quad \mbox{such that} \quad \pi^\ast g_3 = \overline{g}\,.$$
\end{theorem}

$\overline{g}$ is then said to be a {\em projectable spatial metric}.

\paragraph{Proof: }
Given two vector fields, $\vec{v}$ and $\vec{w}$, on ${\cal E}_3$, let $V$ and $W$ be the associated spatial vector fields on ${\cal V}_4$ ---see Proposition \ref{p2}. 
For any $p\in {\cal E}_3$ we can define:
\begin{equation}
\left.g_3(\vec{v},\vec{w})\right|_p := \left.\overline{g}(V,W)\right|_{\pi^{-1}p}
\label{e2.12}
\end{equation}
In order that the latter definition be consistent, it is enough to proof that the right hand side is constant along the worldline $\pi^{-1}p$. Indeed, since $[u,V]=[u,W]=0$, we have that:
$$ u\left\{\overline{g}(V,W)\right\}= {\cal L}(u)\overline{g}(V,W) =0 $$
Finally, the non-degeneracy and positiveness of the metric $g_3$ so defined follow straightforward from Lemma \ref{l1}. \hfill $\Box$

\subsubsection{The contravariant spatial metric\label{SSS2.3.a}}
The spatial metric $\overline{g}$ allows to define a linear map from spatial vectors onto spatial 1-forms:
$$ \left.\begin{array}{lcrlcl}
\forall x\in{\cal V}_4 &\qquad &\overline{g}_x:&{\rm Ker}\,\theta_x&\longrightarrow &[u_x]^\perp\subset T_x^\ast{\cal V}_4 \\
                     &      &       & V_x &\longrightarrow& \overline{g}_x(V_x,\_ )
\end{array}
\right\}    $$
This linear map is invertible. Indeed, to find an inverse image for $\lambda_x\in [u_x]^\perp$ means to solve for $v_x$ the linear system:
\begin{equation}
\overline{g}_x(v_x,w_x)=\langle\lambda_x,w_x\rangle \,, \qquad \forall w_x\in T_x{\cal V}_4 \,.
\label{e2.12a}
\end{equation}
By (\ref{e2.10}), this system is not Cramer's, but it is compatible because $\lambda_x$ vanishes in ${\rm Rad}[\overline{g}]$. Therefore, it is also underdetermined. Given a particular solution $v_{0x}\in T_x{\cal V}_4$ of (\ref{e2.12a}), the sought spatial vector is:
$$ V_x = v_{0x} +\langle\theta_x,v_{0x}\rangle\,u_x $$
where the condition $V_x \in{\rm Ker}\,\theta_x$ has been taken into account.

We shall denote by $\overline{h}$ the inverse map
$$ \lambda\in [u]^\perp \longrightarrow\overline{h}(\lambda)\in {\rm Ker}\,\theta \qquad \mbox{with} \qquad \overline{g}(\overline{h}(\lambda),\_ ) = \lambda $$
The same symbol will be also used to denote the contravariant spatial metric:
$$ \overline{h}: [u]^\perp \times [u]^\perp \longrightarrow R \,,\qquad \mbox{with} \qquad \overline{h}(\lambda,\mu) = \overline{g}(\overline{h}(\lambda),\overline{h}(\mu)) $$

\subsection{Fluid reference frames \label{SS2.4}}
In case that there is not a Riemannian metric on ${\cal E}_3$, we shall relax a little bit the definition.

\begin{definition} \label{d2}
A {\em fluid reference frame} is defined by a triple $({\cal E}_3,\theta,\overline{g})$ ---a timelike 3-congruence, a proper time scale and a spatial metric on ${\cal V}_4$.
\end{definition}

We shall introduce the covariant tensor
\begin{equation}
S:={\cal L}(u) \overline{g} \,.
\label{e2.16}
\end{equation}
In case that $S=0$, the metric $\overline{g}$ is projectable and, by theorem \ref{t1}, a Riemannian metric $g_3$ on ${\cal E}_3$ can be obtained, so that $({\cal E}_3,\theta,g_3)$ is a rigid reference frame.

In general, neither a projected metric on ${\cal E}_3$ does exist, nor the associated Riemannian connexion. Nevertheless, a linear connexion can be set on ${\cal V}_4$ that is specially suited for a given reference frame and, as it will be proved in Theorem \ref{t3}, is closely related to the notion of a Riemannian connexion in ${\cal E}_3$.

\subsubsection{The frame connexion \label{SS2.4.a}}
\begin{theorem} \label{t2}
Let $({\cal E}_3,\theta,\overline{g})$ be a fluid reference frame in a spacetime ${\cal V}_4$. There is a unique symmetric spacetime connexion $\overline{\nabla}$ such that:
\begin{eqnarray}
(a) &\hspace*{3em}& \overline{\nabla} u =0 \label{e2.17a} \\
(b) &\hspace*{3em}& \left(\overline{\nabla}\theta - \frac12 d\theta\right)_\perp =0   \label{e2.17b} \\
(c) &\hspace*{3em}& \left(\overline{\nabla}\overline{g}\right)_\perp =0     \label{e2.17c}
\end{eqnarray}
\end{theorem} 
The subscript $\perp$ after a tensor means its spatial component, in the terms defined in eqs. (\ref{e2.5}), (\ref{e2.7}) and (\ref{e2.7a}). Also notice that the right hand side in equation (\ref{e2.17b}) is the symmetric part of $\overline{\nabla}\theta$.

The proof of the theorem needs to previously prove the following

\begin{lemma} \label{p6}
Let $T$ and $u$ be respectively a tensor and a vector field on ${\cal V}_4$, and $\overline{\nabla}$ a spacetime symmetric connexion such that $\overline{\nabla} u = 0$, then
\begin{equation}
\overline{\nabla}_u T = {\cal L}(u) T 
\label{e2.14}
\end{equation}
\end{lemma}

Indeed, since $\overline{\nabla}_u$ and ${\cal L}(u)$ are derivations of the tensor algebra, it is enough to prove \cite{Godbi2} that (\ref{e2.14}) holds:
\begin{list}
{(\alph{llista})}{\usecounter{llista}}
\item{for functions, which is obvious,}
\item{for any vector field $v$. In this case we have that
$$ {\cal L}(u) v = [u,v] = \overline{\nabla}_u v - \overline{\nabla}_v u = \overline{\nabla}_u v $$
where the symmetry of $\overline{\nabla}$ has been taken into account. And}
\item{for any 1-form $\alpha\in\Lambda^1{\cal V}_4$. Given any vector field $w$ on ${\cal V}_4$,
\begin{eqnarray*}
\hspace*{3em}\langle\overline{\nabla}_u\alpha,w\rangle &=& u\langle\alpha,w\rangle - \langle\alpha,\overline{\nabla}_u w\rangle \\ 
  & = &  u\langle\alpha,w\rangle - \langle\alpha,{\cal L}(u) w\rangle = \langle{\cal L}(u)\alpha,w\rangle 
\hspace*{6em}\hfill \Box
\end{eqnarray*}}
\end{list}
\paragraph{Proof of theorem \ref{t2}:}
Let $v$ and $w$ be two vector fields on ${\cal V}_4$. According to (\ref{e2.5}) we can split them in their spatial and time parts:
\begin{equation}
 v = V - \langle\theta,v\rangle \, u \qquad    w= W - \langle\theta,w\rangle \, u \qquad{\rm with} \qquad \langle\theta,V\rangle =\langle\theta,W\rangle =0 
\label{e2.14a}
\end{equation}
So that, taking Lemma \ref{p6} into account, we can write:
$$ \overline{\nabla}_vw = \overline{\nabla}_V W - V\langle\theta,w\rangle\,u - \langle\theta,w\rangle {\cal L}(u) w $$
Hence, only the first term in the right hand side has to be determined and. By (\ref{e2.5}) again  it can be split as: 
$$ \overline{\nabla}_V W = (\overline{\nabla}_V W)_\perp - \langle\theta,\overline{\nabla}_V W\rangle\,u  $$
The second term on the right hand side can be derived from (\ref{e2.17b}). Indeed, taking  (\ref{e2.17b}) and (\ref{e2.14a}) into account we readily obtain:
$$\langle\theta,\overline{\nabla}_V W\rangle = -\langle\overline{\nabla}_V\theta, W\rangle = -\frac12 d\theta(V,W) = \frac12 \langle\theta, [V,W]\rangle \,.$$  
While the spatial part can be derived from (\ref{e2.17c}) by applying a similar algorithm  as in the case of Riemannian connexions. It results that: \hspace*{.4em} $\forall T\in[\theta]^\perp$
\begin{eqnarray*}
2\, \overline{g} (\overline{\nabla}_V W,T)&=& V\overline{g}(W,T) + W\overline{g}(V,T) - T\overline{g}(V,W) \\
  & & + \overline{g}([V,W],T) + \overline{g}([T,V],W) + \overline{g}([T,W],V)
\end{eqnarray*}
that determines completely $\overline{\nabla}_V W$ because, by Lemma \ref{l1}, $\overline{g}$ restricted to ${\rm Ker}\,\theta$ is nondegenerate. \hfill $\Box$

An interesting consequence can also be drawn from condition (\ref{e2.17b}). Indeed, since $\overline{\nabla}$ is a symmetric connexion, $\overline{\nabla} \theta -\frac12 d\theta$ is a symmetric tensor whose purely spatial component vanishes. We can therefore write:
$$\overline{\nabla}\theta - \frac12\,d\theta = \theta\otimes \alpha + \alpha\otimes\theta \qquad\mbox{for some 1-form}\quad \alpha$$ 
By contraction with $u$ and using Lemma \ref{l2}, we obtain that:
$$ -\alpha + \langle\alpha,u\rangle \theta = \overline{\nabla}_u\theta -\frac12 i(u)d\theta =\frac12{\cal L}(u)\theta$$ 
which immediately leads to $\alpha=-\frac12\,{\cal L}(u)\theta$. We have therefore that:
\begin{equation}
\overline{\nabla}\theta = \frac12\,d\theta - \frac12\, \left(\theta\otimes {\cal L}(u)\theta + {\cal L}(u) \theta \otimes \theta\right)
\label{e2.14b}
\end{equation}
We then split $d\theta$ in a spatial part and a time part ---as indicated in section \ref{SS2.2}--- and arrive at:
$$ d\theta = \Theta - (\theta\otimes {\cal L}(u)\theta - {\cal L}(u) \theta \otimes \theta)$$ 
where $\Theta$ is the vorticity of the congruence and $i(u)\Theta =0$. Now, using together the last two equations we finally have
\begin{equation}
\overline{\nabla} \theta = \frac12 \Theta - \theta\otimes {\cal L}(u)\theta
\label{e2.14c}
\end{equation}

For further developments it will be useful to have an expression for the connexion coefficients $\overline{gamma}^\alpha_{\mu\nu}$ for $\overline{\nabla}$. We shall derive an expression for them referred to the coefficients $\gamma^\alpha_{\mu\nu}$ of the spacetime Riemannian connexion in ${\cal V}_4$, in terms of the difference tensor:
\begin{equation}
B(v,w) := \overline{\nabla}_v w -\nabla_v w \,, \qquad v, w \quad \mbox{vector fields on }{\cal V}_4 
\label{e2.18}
\end{equation}
whose components in a basis are:
\begin{equation}
B^\alpha_{\mu\nu} = \overline{gamma}^\alpha_{\mu\nu} - \gamma^\alpha_{\mu\nu}
\label{e2.19}
\end{equation}
Since both, $\overline{\nabla}$ and $\nabla$, are symmetric connexions, $B$ is a symmetric tensor: \hspace*{1em} $B^\alpha_{\mu\nu}=B^\alpha_{\nu\mu}$.

Notice right away that, as a consequence of (\ref{e2.17a}), $\overline{\nabla}_u \overline{g} = {\cal L}(u)\overline{g} = S$.

Furthermore, recalling that $\overline{g}$ is a spatial metric and taking (\ref{e2.17a}) into account, we have that, for any two vector fields $t$ and $w$ on ${\cal V}_4$:
$$ \overline{\nabla}_t\overline{g}(u,w) = t\left(\overline{g}(u,w)\right) - \overline{g}(\overline{\nabla}_t u,w) - \overline{g} (u,\overline{\nabla}_t w) = 0  \,. $$
From the latter and from eq. (\ref{e2.17c}) it follows immediately that, $\forall t,v,w$ fields on ${\cal V}_4$:
\begin{equation}
\overline{\nabla}_t \overline{g} (v,w) = - \langle\theta, t\rangle\,S(v,w)
\label{e2.20}
\end{equation} 

When expressed in a basis, equations (\ref{e2.17a}), (\ref{e2.17b}) and (\ref{e2.20}) respectively yield:
\begin{eqnarray}
 & & \nabla_\mu u^\alpha + B^\alpha_{\mu\nu} u^\nu = 0             \label{e2.21a} \\
 & & P_\mu^\rho \, P_\nu^\sigma \left( \nabla_\rho \theta_\sigma + \nabla_\sigma \theta_\rho - 2  B^\alpha_{\rho\sigma} \theta_\alpha \right) = 0                      \label{e2.21b} \\
 & & \nabla_\lambda \overline{g}_{\rho\sigma} -B^\tau_{\lambda\rho}\overline{g}_{\tau\sigma} -B^\tau_{\lambda\sigma} \overline{g}_{\rho\tau} = - \theta_\lambda S_{\rho\sigma}            \label{e2.21c} 
\end{eqnarray}
where $P_\mu^\rho$ is the spatial projector (\ref{e2.8}).

This system of equations can be solved for $B^\alpha_{\mu\nu}$ and leads to:
\begin{equation}
B^\alpha_{\mu\nu} = \frac12 \,\overline{h}^{\alpha\rho} \left(\nabla_\mu\overline{g}_{\rho\nu}+\nabla_\nu\overline{g}_{\rho\mu}-\nabla_\rho\overline{g}_{\mu\nu} + \theta_\mu S_{\rho\nu} + \theta_\nu S_{\rho\mu} -\theta_\rho S_{\mu\nu}\right) - \frac12 u^\alpha \Sigma_{\mu\nu}
\label{e2.22}
\end{equation}
with
$$ \Sigma_{\mu\nu} :=P_\mu^\rho \, P_\nu^\sigma \left( \nabla_\rho \theta_\sigma + \nabla_\sigma\theta_\rho \right)   $$

For further developments, it will be useful the covariant tensor:
\begin{equation}
B_{\mu\nu\|\rho} := \overline{g}_{\rho\alpha} B^\alpha_{\mu\nu}
\label{e2.23a}
\end{equation}
which can be written as 
\begin{equation}
B_{\mu\nu\|\rho} = \frac12 \left( \nabla_\mu\overline{g}_{\rho\nu}+\nabla_\nu\overline{g}_{\rho\mu}-\nabla_\rho\overline{g}_{\mu\nu} + \theta_\mu S_{\rho\nu} + \theta_\nu S_{\rho\mu} -\theta_\rho S_{\mu\nu} \right) \,,
\label{e2.23b}
\end{equation}
as it easily follows from (\ref{e2.21c}).

\subsubsection{Curvature tensors \label{SSS2.4.b}}
The curvature tensor for the connexion $\overline{\nabla}$ is:
\begin{equation}
\overline{R}(v,w)z=\overline{\nabla}_v\overline{\nabla}_w z -\overline{\nabla}_w\overline{\nabla}_v z-\overline{\nabla}_{[v,w]} z
\label{e2.24}
\end{equation}
for any three vector fields $v$, $w$ and $z$ tangent to ${\cal V}_4$. As any curvature tensor, $\overline{R}$ is skewsymmetric in the variables $v$ and $w$ and the first Bianchi identity holds:
\begin{equation}
\overline{R}(v,w)z + \overline{R}(w,z) v + \overline{R}(z,v)w =0
\label{e2.25}
\end{equation}
Also, as an immediate consequence of (\ref{e2.17a}) and (\ref{e2.24}), it follows that:
\begin{equation}
\overline{R}(v,w) u = 0  \qquad \qquad \forall v, w
\label{e2.26}
\end{equation}

$\overline{R}(v,w)z$ is a vector in the spacetime and can be separated in a spatial and a time part, as in eq. (\ref{e2.5}). The coefficient of $u$ in the time part is 
\begin{equation}
\langle\theta,\overline{R}(v,w)z\rangle = \langle \overline{\nabla}_v\overline{\nabla}_w \theta -\overline{\nabla}_w\overline{\nabla}_v \theta -\overline{\nabla}_{[v,w]} \theta, z \rangle 
\label{e2.27}
\end{equation}
whereas the spatial part is fully determined by the covariant curvature tensor
\begin{equation}
\overline{K}(t,z,v,w) :=\overline{g}(t,\overline{R}(v,w)z)
\label{e2.28}
\end{equation}
[The latter plays in our construct a similar role as the Riemann tensor does in Riemannian manifolds, except for that the metric $\overline{g}$ is degenerate, and the contraction in the right hand side erases any information concerning the time component (\ref{e2.27}).]

After a short calculation, from (\ref{e2.17c}) we obtain that:
\begin{equation}
\overline{K}(t,z,v,w) + \overline{K}(z,t,v,w) = d\theta (v,w)\,S(t,z) + \langle\theta,w\rangle\,\overline{\nabla}_v S(t,z) - \langle\theta,v\rangle\,\overline{\nabla}_w S(t,z)
\label{e2.29}
\end{equation}

Therefore, the covariant curvature tensor is not skewsymmetric in the first pair of indices, in contrast with the Riemann curvature tensor, except in the case $S=0$, i. e., when the metric $\overline{g}$ is projectable. 

\subsubsection{Components and symmetries \label{SSS2.4.c}}
The difference between the  components of the curvature tensors, $\overline{R}^\mu_{\;\nu\alpha\beta }$ and $R^\mu_{\;\nu\alpha\beta }$, respectively associated to the connexions $\overline{\nabla}$ and $\nabla$, depends on the difference tensor:
\begin{equation}
\overline{R}^\mu_{\;\nu\alpha\beta } = R^\mu_{\;\nu\alpha\beta } + 2\,\nabla_{[\alpha}B_{\beta ]\nu}^\mu + 2\,B^\mu_{\rho[\alpha}B_{\beta ]\nu}^\rho
\label{e2.201}
\end{equation}
[The above expression follows immediately from writing each curvature tensor in terms of the respective connexion coefficients\cite{CHOQ1}, using the relation (\ref{e2.19}).]

As for the covariant curvature tensor, the definition (\ref{e2.28}) can be written in terms of components as
$$ \overline{K}_{\mu\nu\alpha\beta } :=\overline{g}_{\mu\rho}\overline{R}^\rho_{\;\nu\alpha\beta } $$
which, combined with (\ref{e2.201}) and taking (\ref{e2.22}) and (\ref{e2.23a}) into account, leads to
\begin{equation}
\overline{K}_{\mu\nu\alpha\beta} = \overline{g}_{\mu\rho} R^\rho_{\;\nu\alpha\beta } + 2\,\nabla_{[\alpha}B_{\beta ]\nu\|\mu} + 2\,B^\rho_{\nu[\alpha}B^{\,}_{\beta ]\mu\|\rho} - S_{\mu\rho} B^\rho_{\nu[\alpha} \theta_{\beta ]}
\label{e2.202}
\end{equation}

In terms of components, the symmetries of $\overline{K}$ commented above yield:
\begin{eqnarray}
& & \overline{K}_{\mu\nu\alpha\beta } = -\overline{K}_{\mu\nu\beta \alpha} \label {e2.203} \\
& & \overline{K}_{\mu\nu\alpha\beta } + \overline{K}_{\nu\mu\alpha\beta } = 2 \nabla_{[\alpha}\theta_{\beta]} S_{\mu\nu} - 2 \theta_{[\alpha}\overline{\nabla}_{\beta]} S_{\mu\nu}   \label{e2.204}
\end{eqnarray}
and from (\ref{e2.26}) it follows that:
$$ \overline{K}_{\mu\nu\alpha\beta} u^\nu =\overline{K}_{\mu\nu\alpha\beta} u^\mu =0 \,.$$
The first Bianchi identity (\ref{e2.25}) implies that
\begin{equation}
\overline{K}_{\mu\nu\alpha\beta} + \overline{K}_{\mu\alpha\beta\nu} + \overline{K}_{\mu\beta\nu\alpha} =0
\label{e2.205}
\end{equation}

These symmetry and orthogonality relations are the starting point to derive a spatial tensor (i. e., orthogonal to $u^\alpha$ in any index) $T_{\mu\nu\alpha\beta }$ that has precisely the same kind of symmetries as a Riemann tensor. 
A painful calculation, whose details will not be explicited here, leads to:
\begin{equation}
T_{\mu\nu\alpha\beta } := \overline{K}_{\mu\nu\alpha\beta } - Q_{\mu\nu\alpha\beta } + \theta_{[\alpha}L_{\beta ]\mu\nu}
\label{e2.208}
\end{equation}
with
\begin{equation}
\left. \begin{array}{l}
Q_{\mu\nu\alpha\beta}:= \frac12 \left(S_{\mu\nu}\Theta_{\alpha\beta } + S_{\nu[\alpha}\Theta_{\beta ]\mu} - S_{\mu[\alpha}\Theta_{\beta ]\nu} \right) \\
L_{\beta \mu\nu} := H_{\beta \mu\nu} +H_{\nu\mu\beta } -H_{\mu\nu\beta }  \\
H_{\beta \mu\nu} := P_\beta ^\sigma \overline{\nabla}_\sigma S_{\mu\nu} + b_\beta S_{\mu\nu}
\end{array}  \right\}
\label{e2.209}
\end{equation}
where the expressions: $\Theta_{\alpha\beta }= 2 P^\rho_\alpha P^\sigma_\beta \,\nabla_{[\rho} \theta_{\sigma]}$ for the spatial part of $d\theta$ and $b_\beta := u^\rho \nabla_\rho \theta_\beta$ have been used used.

Notice that $ L_{\beta \mu\nu} = 2 u^\alpha \overline{K}_{\mu\nu\alpha\beta }  $
is also orthogonal to $u^\rho$ for all indices and presents the symmetry: $L_{\beta \mu\nu}=L_{\nu\mu\beta } \,.$

Moreover, it follows immediately from (\ref{e2.209}) that
\begin{equation}
L_{\beta (\mu\nu)} = P_\beta ^\sigma \overline{\nabla}_\sigma S_{\mu\nu} + b_\beta S_{\mu\nu}
\label{e2.212}
\end{equation}

On its turn, the second Bianchi identity is
\begin{equation}
\eta^{\sigma\lambda\alpha\beta } \overline{\nabla}_\lambda \overline{R}^\mu_{\;\nu\alpha\beta } =0
\label{e2.206}
\end{equation}
where $\eta^{\sigma\lambda\alpha\beta }$ is any completely skewsymmetric tensor, e. g., the contravariant volume tensor for the Riemannian metric. In terms of the covariant curvature tensor, the latter identity yields:
\begin{equation}
\eta^{\sigma\lambda\alpha\beta } \left(\overline{\nabla}_\lambda \overline{K}_{\mu\nu\alpha\beta } + \theta_\lambda \overline{R}^\rho_{\;\nu\alpha\beta } S_{\mu\rho} \right) = 0
\label{e2.207}
\end{equation}

\subsubsection{Projectable metrics \label{SSS2.4.d}}
If the metric $\overline{g}$ is projectable in the sense of Theorem \ref{t1}, it exists a Riemannian metric on ${\cal E}_3$. Then the following question arises: is there any relationship between the connexion $\overline{\nabla}$ on ${\cal V}_4$ and the Riemannian connexion $\stackrel{3}{\nabla}$ in ${\cal E}_3$? Theorem \ref{t3} provides a positive answer to this question, and Propositions \ref{p7} to \ref{p9} are preliminar to its proof.

\begin{proposition} \label{p7}
Let $\overline{g}$ be a projectable spatial metric, then the covariant curvature tensor $\overline{K}$ has the same symmetries as a Riemann tensor:
\begin{eqnarray}
{\rm (a)} & \qquad & \overline{K}(t,z,v,w) + \overline{K}(t,v,w,z) +\overline{K}(t,w,z,v) = 0 \label{e2.30a} \\
{\rm (b)} & \qquad & \overline{K}(t,z,v,w) + \overline{K}(z,t,v,w) = 0   \label{e2.30b}\\
{\rm (c)} & \qquad & \overline{K}(t,z,v,w) - \overline{K}(v,w,t,z) = 0   \label{e2.30c}
\end{eqnarray}
\end{proposition}

\paragraph{Proof: }
(a) results from the Bianchi identity (\ref{e2.25}). (b) follows on substituting $S=0$ ---recall that $\overline{g}$ is projectable--- into eq (\ref{e2.29}). Finally, (c) results from recombining four identities obtained by cyclic permutation of $t,z,v,w$ in (\ref{e2.30a}) with the help of (\ref{e2.30b}). \hfill $\Box$

\begin{proposition} \label{p8}
If $\overline{g}$ is projectable, then $\pi_\ast[\overline{R}(u,v)w]=0$.
\end{proposition}

Indeed, using (\ref{e2.30c}) and taking the definition \ref{d0} into account,  we have that, $\;\forall t\in T{\cal V}_4\,,$
$$ \overline{g}(t,\overline{R}(u,v)w) = \overline{K}(t,w,u,v) = \overline{K}(u,v,t,w) = \overline{g}(u,\overline{R}(t,w)v) =0  $$
Therefore, $\overline{R}(u,v)w \in {\rm Rad}\overline{g} ={\rm span}\,[u]$, whence: $\pi_\ast[\overline{R}(u,v)w]=0$ \hfill $\Box$

\begin{proposition} \label{p9}
Let $\overline{g}$ be projectable and let $v,w$ be two projectable vector fields, then $\overline{\nabla}_v w$ is projectable. Moreover, if $\pi_\ast v =\pi_\ast v^\prime$ and $\pi_\ast w =\pi_\ast w^\prime$, then $\pi_\ast(\overline{\nabla}_v w) = \pi_\ast(\overline{\nabla}_{v^\prime} w^\prime)$.
\end{proposition}

\paragraph{Proof: }
From (\ref{e2.14}) we have that:
$${\cal L}(u)\overline{\nabla}_v w = \overline{\nabla}_u\overline{\nabla}_v w = \overline{\nabla}_v\overline{\nabla}_u w + \overline{\nabla}_{[u,v]} w +\overline{R}(u,v)w $$
Since $v$ and $w$ are projectable, $\exists f_1,f_2 \in\Lambda^0{\cal V}_4$ such that:
$$ [u,v] = f_1 u \qquad {\rm and} \qquad \overline{\nabla}_u w =[u,w] = f_2 u $$
and, since $\overline{g}$ is projectable, from Proposition \ref{p8} we have that:
$$ \exists f_3 \in\Lambda^0{\cal V}_4 \qquad \mbox{such that} \qquad \overline{R}(u,v)w =f_3 u$$
Taking now into account that $\overline{\nabla}_v u =0$ we arrive at
$$ {\cal L}(u)\overline{\nabla}_v w  = \left\{v(f_2) + f_1 f_2 + f_3\right\} u  $$
whence, by Proposition \ref{p1}, it follows that $\overline{\nabla}_v w$ is projectable.

As for the second statement, if $\pi_\ast v =\pi_\ast v^\prime$ and $\pi_\ast w =\pi_\ast w^\prime$, then $ \exists f, l\in \Lambda^0{\cal V}_4$ such that
$$ v^\prime = f u + v \qquad {\rm and} \qquad w^\prime = l u + w $$
Thus, using (\ref{e2.17a}) and that $w$ is projectable, we can write
$$ \overline{\nabla}_{v^\prime} w^\prime = \overline{\nabla}_v w+ [f f_2+v^\prime(l)]\,u $$
and the proof follows immediately. \hfill $\Box$

\begin{theorem} \label{t3}
Let $\vec{v},\, \vec{w}$ be two vector fields on ${\cal E}_3$ and $v,\,w$ any two vector fields on ${\cal V}_4$, such that: $\pi_\ast v =\vec{v}$ and $\pi_\ast w =\vec{w}$, then:
$\pi_\ast(\overline{\nabla}_v w) =\stackrel{3}{\nabla}_{\vec{v}} \vec{w} $
\end{theorem} 

\paragraph{Proof: }
Given $\vec{v}$ and $\vec{w}$, let us define:
$$ \hat\nabla_{\vec{v}} \vec{w} := \pi_\ast\left(\overline{\nabla}_v w\right) $$
It follows from Proposition \ref{p9} that the definition is unambiguous. Moreover, it is easy to see that $\hat\nabla$ is a symmetric connexion on ${\cal E}_3$.

Now consider $\vec{v}_p,\, \vec{w}_p\in T{\cal E}_3$ and let $\vec{v},\, \vec{w}$ be two vector fields on ${\cal E}_3$ such that: {\em (a)} they take the prescribed values at $p$ and {\em (b)} $\hat\nabla_{\vec{t}} \vec{v} = \hat\nabla_{\vec{t}} \vec{w} =0$, for some $\vec{t}$. We then choose $t\,, v\,,w\,$, vector fields on ${\cal V}_4$ such that $\pi_\ast t =\vec{t}$, $\pi_\ast v =\vec{v}$ and $\pi_\ast w =\vec{w}$ (see Proposition \ref{p9}). Then
$$
\left.\hat\nabla_{\vec{t}} g_3(\vec{v},\vec{w})\right|_p = \vec{t}\left(g_3(\vec{v},\vec{w})\right)_p - g_3(\hat\nabla_{\vec{t}}\vec{v},\vec{w})_p - g_3(\vec{v},\hat\nabla_{\vec{t}}\vec{w})_p 
 = t\left(\overline{g}(v,w)\right)_p  
$$
Taking now into account (\ref{e2.20}) and the fact that $S=0$ implies $\overline{\nabla}_t\overline{g}=0$, we can write:
$$ t\left[\overline{g}(v,w)\right]_p= \overline{g}(\overline{\nabla}_t v,w)_{\pi^{-1}p} + \overline{g}(v,\overline{\nabla}_t w)_{\pi^{-1}p} 
$$
that, since $\overline{g}=\pi^\ast g_3$ (Theorem \ref{t1}), is equal to
$$g_3(\hat\nabla_{\vec{t}}\vec{v},\vec{w})_p + g_3(\vec{v},\hat\nabla_{\vec{t}}\vec{w})_p = 0 $$
Therefore, $\hat\nabla g_3 =0$ and, from the uniqueness of the Riemannian connexion associated to $g_3$, it follows that $\hat\nabla = \stackrel{3}{\nabla}$. \hfill $\Box$ 

The next result, which follows immediatelly and we shall state without proof, relates the Riemann tensor for $\stackrel{3}{\nabla}$ and the covariant curvature tensor for $\overline{\nabla}$.

\begin{corollary} \label{c3}
Let $\stackrel{3}R$ be the Riemann tensor for the space Riemannian connexion $\stackrel{3}{\nabla}$ and $\overline{K}$ the covariant curvature tensor for $\overline{\nabla}$. Then:
$$ \overline{K} = \pi^\ast \stackrel{3}R $$
\end{corollary}

Later on we shall be interested in a spatial metric $\overline{g}$ that is rigid and has {\em constant curvature}. The latter will mean that the Riemannian metric $g_3$ that results from projecting $\overline{g}$ onto ${\cal E}_3$ has constant curvature.

\subsection{Subframes of reference \label{SS2.5}}
Let $({\cal E}_3,\overline{g},\theta)$ be a reference frame on ${\cal V}_4$ and let ${\cal V}_3$ be a 3-submanifold of ${\cal V}_4$ which is also tangent to $u$ at every point: $\forall p\in{\cal V}_3\,, \; u_p\in T_p{\cal V}_3$, and let $j:{\cal V}_3\longrightarrow {\cal V}_4$ be the submanifold embedding. Then ${\cal E}_3$ induces a timelike 2-congruence, ${\cal E}_2$ in ${\cal V}_3$. Furthermore, 
\begin{equation} 
\tilde{\theta}:= j^\ast \theta    \qquad \mbox{and} \qquad  \tilde{g} := j^\ast \overline{g}
\label{2.32}
\end{equation}
are respectively a time scale and a spatial metric on ${\cal V}_3$\footnote{The validity of the equalities along this subsection will be tacitly restricted to points $p\in{\cal V}_3$, unless the contrary is explicitly stated}. 

Thus $({\cal E}_2,\tilde{g},\tilde{\theta})$ is a reference frame on ${\cal V}_3$ and a {\it subframe of reference} of $({\cal E}_3,\overline{g},\theta)$.

The submanifold ${\cal V}_3$ selects a class of orthogonal 1-forms, $\zeta\in \Lambda^1{\cal V}_4$ such that:
\begin{equation}
j^\ast\zeta = 0
\label{e2.33}
\end{equation}
The latter only determines $\zeta_p$ at any point $p\in{\cal V}_3$ except for a factor. We can thus complete $\zeta$ on ${\cal V}_4$ so that it is spatial: $\langle\zeta,u\rangle =0$

Since $u_p\in T_p{\cal V}_3$, $\,\forall p\in {\cal V}_3\,$, then $\langle\zeta,u\rangle_p=0$. Hence, $\zeta_p$ is a spatial 1-form and it can be applied the contravariant metric $\overline{h}$ ---subection \ref{SSS2.3.a}. Let us choose $\zeta$ so that
\begin{equation}
\overline{h}(\zeta,\zeta) = +1
\label{e2.34}
\end{equation} 
and denote by $n\in [\theta]^\perp $ the spatial vector defined on ${\cal V}_4$ as
\begin{equation}
n := \overline{h}(\zeta,\_ ) \qquad {\rm or} \qquad  \zeta = \overline{g}(n,\_ )
\label{e2.35}
\end{equation}
which is obviously $\overline{g}$-unitary: $\;\overline{g}(n,n)=+1$\,.

\begin{definition} \label{d3}
The 2-subframe $({\cal E}_2,\tilde{g},\tilde{\theta})$ is said to be {\em rigidly embedded} in the 3-frame $({\cal E}_3,\overline{g},\theta)$ if $\left.S(n,\_\,)\right|_p =0$\,, for all $p\in{\cal V}_3$\,.
\end{definition}

\subsubsection{Projection of $\overline{\nabla}$ onto the 2-congruence \label{SSS2.5.1}}
Let $v$ and $w$ be two vector fields on ${\cal V}_3$, then $\overline{\nabla}_v w$ is not necessarily tangent to ${\cal V}_3$, but can be split into its tangent and an orthogonal components relatively to ${\cal V}_3$ :
\begin{equation}
\overline{\nabla}_v w = \tilde{\nabla}_v w + \langle\zeta,\overline{\nabla}_v w\rangle \,n
\label{e2.37}
\end{equation}
We shall now see that in some interesting cases $\tilde{\nabla}$ defines a connexion on ${\cal V}_3$.

\begin{lemma} \label{l2}
If $({\cal E}_2,\tilde{g},\tilde{\theta})$ is rigidly embedded in $({\cal E}_3,\overline{g},\theta)$, then $\overline{\nabla}_v n$ is tangent to ${\cal V}_3$, for any $v$ tangent to ${\cal V}_3$.
\end{lemma}

Indeed, from (\ref{e2.35}) we have that
$$ \langle\zeta,\overline{\nabla}_v n\rangle = \overline{g}(n,\overline{\nabla}_v n) = - \frac12 \overline{\nabla}_v\overline{g}(n,n) $$
which, using (\ref{e2.20}) and the condition of rigid embedding, yields:
$$\hspace*{5em} \langle\zeta,\overline{\nabla}_v n\rangle = \frac12 \langle\theta,v\rangle\,S(n,n)=0 \qquad\hspace*{5em}
\hfill \Box$$

The {\it second fundamental form} of the 2-subframe is the covariant tensor on $T{\cal V}_3$ defined by:
\begin{equation}
\phi(v,w) := \overline{g}(\overline{\nabla}_v n,w) \qquad \qquad v,\,w \in T{\cal V}_3 
\label{e2.39}
\end{equation}

\begin{proposition}  \label{p10}
If the embedding is rigid and $v, w \in T{\cal V}_3$, then:
\begin{equation}
\phi(v,w) = - \langle \zeta, \overline{\nabla}_v w\rangle 
\label{e2.40}
\end{equation}
and $\phi$ is symmetric
\end{proposition}

Indeed, 
$$\phi(v,w) = \overline{g}(\overline{\nabla}_v n,w) = v\langle\zeta,w\rangle - \overline{\nabla}_v\overline{g}(n,w) - \langle\zeta,\overline{\nabla}_v w\rangle $$
Now, taking into account that $\langle\zeta, w\rangle =0$, equation (\ref{e2.20}) and Definition \ref{d3}, the expression (\ref{e2.40}) readily follows. The symmetry is then obvious because $\overline{\nabla}_v w - \overline{\nabla}_w v = [v,w]$ is tangent to ${\cal V}_3$. \hfill $\Box$

The tangential component of $\overline{\nabla}_v w$ can also be written as:
\begin{equation}
\tilde{\nabla}_v w = \overline{\nabla}_v w + \phi(v,w)\,n
\label{e2.42}
\end{equation}

\begin{theorem}  \label{t4}
$\tilde{\nabla}$ is the frame connexion of the 2-subframe $({\cal E}_2,\tilde{g},\tilde{\theta})$
\end{theorem}

\paragraph{Proof:}
We have to prove that $\tilde{\nabla}$ is symmetric and meets the three conditions (a), (b) and (c) listed in Theorem \ref{t2}. Let us first consider the difference:
\begin{eqnarray*}
\tilde{\nabla}_v w -\tilde{\nabla}_w v &=& \overline{\nabla}_v w - \overline{\nabla}_w v - \langle\zeta, \overline{\nabla}_v w - \overline{\nabla}_w v\rangle  \\
  & = & [v,w] - \langle\zeta, [v,w]\rangle = [v,w] 
\end{eqnarray*}
because, being ${\cal V}_3$ a submanifold, $\langle\zeta, [v,w]\rangle=0$. Hence $\tilde{\nabla}$ is symmetric. 

Condition (a) follows immediately from the fact that $\tilde{\nabla}_v u$ is the projection of $\overline{\nabla}_v u =0$.

Now we also have that $\forall t,v,w \in T {\cal V}_3$, 
\begin{eqnarray*}
\tilde{\nabla}_v\tilde{g}(w,t) &=& v\left(\tilde{g}(w,t)\right)-\tilde{g}(\tilde{\nabla}_v w,t) -\tilde{g}(w,\tilde{\nabla}_v t) \\
 & = & v\left(\overline{g}(w,t)\right)-\overline{g}(\overline{\nabla}_v w,t) -\overline{g}(w,\overline{\nabla}_v t) = \overline{\nabla}_v\overline{g}(w,t)
\end{eqnarray*}
where (\ref{e2.42}) and the fact that $\overline{g}(n,w) = \overline{g}(n,t) =0$ have been taken into account. Using now (\ref{e2.20}) we arrive at
$$ \tilde{\nabla}_v\tilde{g}(w,t) = - \langle\theta,v\rangle\,{\cal L}(u)\overline{g}(w,t) = - \langle\tilde{\theta},v\rangle\,{\cal L}(u)\tilde{g}(w,t)=- \langle\tilde{\theta},v\rangle\,\tilde{S}(w,t) \,,$$ which proves that condition (c) is met.

Finally, denoting the differential operator on $\Lambda{\cal V}_3$ by $\tilde d$, we have that, $\,\forall v,w \in T{\cal V}_3\cap {\rm Ker\,}\tilde\theta$,
\begin{eqnarray}
\left(\tilde{\nabla}\tilde{\theta} -\frac12\,\tilde d\tilde{\theta} \right)(w,v) &=& \frac12\,\langle\tilde{\nabla}_w \theta, v\rangle + \frac12\,\langle\tilde{\nabla}_v \theta, w\rangle \nonumber  \\
&=& \frac12\,w\langle\tilde{\theta},v\rangle - \frac12\langle\tilde{\theta},\tilde{\nabla}_w v\rangle + \frac12\, v\langle\tilde{\theta},w\rangle - \frac12\langle\tilde{\theta},\tilde{\nabla}_v w\rangle
\label{e2.42a}
\end{eqnarray}
Using now (\ref{e2.37}) and the fact that $\langle\theta,n\rangle=0$, we also have that
$$  \langle\tilde{\theta},\tilde{\nabla}_w v\rangle = \langle\theta,\overline{\nabla}_w v\rangle \qquad {\rm and} \qquad 
\langle\tilde{\theta},\tilde{\nabla}_v w\rangle = \langle\theta,\overline{\nabla}_v w\rangle $$
which, substituted into (\ref{e2.42a}) and taking (\ref{e2.17b}) into account, yield
$$ \left(\tilde{\nabla}\tilde{\theta} -\frac12\,\tilde d\tilde{\theta} \right)(w,v) =
\left(\overline{\nabla}\theta -\frac12\,d\theta \right)(w,v) = 0 $$
which proves condition (b). \hfill$\Box$

Later on it will be useful to have at hand the components of $\phi$ in a given basis. Let us consider the vector field on ${\cal V}_4$ that results from ``raising the index'' of $\zeta$ with the spacetime metric: $z^\alpha = g^{\alpha\beta } \zeta_\beta $, and let $\hat{z}$ be the $g$-unitary vector:
$$\hat z^\alpha := \frac1{\|\zeta\|}\, g^{\alpha\beta }  \zeta_\beta \qquad {\rm with} \qquad \|\zeta\| := \sqrt{g^{\mu\nu}\zeta_\mu \zeta_\nu}$$
Thus, $\zeta = g( z,\_ ) = \|\zeta\|\,g(\hat z,\_ ) $\, and, from the expression (\ref{e2.40}) and taking (\ref{e2.19}) into account, we can write:
\begin{eqnarray}
\phi(v,w) &=& -\langle\zeta,\overline{\nabla}_v w\rangle= -\langle\zeta,\nabla_v w + B(v,w)\rangle \nonumber \\
   & =& - \|\zeta\|\,g(\hat z, \nabla_v w) - \langle\zeta, B(v,w)\rangle =  \|\zeta\|\,\Phi(v,w) - \langle\zeta, B(v,w)\rangle   \label{e2.42.2}
\end{eqnarray}
where $\Phi$ is the second fundamental form of ${\cal V}_3$ as a Riemannian submanifold of ${\cal V}_4$.

Expliciting its components, we easily obtain that
$$ \phi_{\mu\nu} = \|\zeta\|\Phi_{\mu\nu} - B_{\mu\nu\|\rho} n^\rho $$
or, in case that the embedding is rigid and using (\ref{e2.23b}), 
\begin{equation}
\phi_{\mu\nu} = \|\zeta\|\Phi_{\mu\nu} + \nabla_{(\mu}\overline{g}_{\nu)\rho} n^\rho - \frac12 \nabla_n \overline{g}_{\mu\nu}
\label{e2.42.3}
\end{equation}

It is interesting to remark that, by definition, $\phi_{\mu\nu} n^\nu =0$. Hence, if we introduce the projector $\Pi^\nu_\mu = \delta^\nu_\mu - n^\nu \zeta_\mu $, we have that:
\begin{equation}
\phi_{\mu\nu} = \Pi^\alpha_\mu\,\Pi^\beta _\nu\,\phi_{\alpha\beta }
\label{e2.42.4}
\end{equation}

We have also that $\phi_{\mu\nu} u^\nu =0$. Indeed, for any $v$ it follows from (\ref{e2.40}) and  (\ref{e2.17a}) that: $ \phi(v,u) = -\langle\zeta, \overline{\nabla}_v u\rangle = 0 \,.$

\subsubsection{$\overline{g}$-Gauss and $\overline{g}$-Codazzi-Mainardi equations \label{SSS2.5.2}}
In a way similar as it is done for Riemannian submanifolds, we shall consider the tangent and normal components of $\overline{R}(v,w)t$ for $v,w,t\in T{\cal V}_3$. Only the case of rigid embedding will be considered from now on. After a short calculation that uses eq. (\ref{e2.42}) we obtain for the tangent part:
\begin{equation}
\left(\overline{R}(v,w)t\right)_\top = \tilde{R}(v,w)t - \phi(v,t)\,\overline{\nabla}_w n + \phi(w,t)\,\overline{\nabla}_v n 
\label{e2.43}
\end{equation}
which is the analogous to the Gauss equation.

As for the normal component, after a little calculation we obtain:
\begin{equation}
\langle\zeta,\overline{R}(v,w)t\rangle = \tilde{\nabla}_v \phi(w,t)-\tilde{\nabla}_w \phi(v,t)
\label{e2.44}
\end{equation}
that has a similar role as the Codazzi-Mainardi equation.

In terms of the covariant curvature tensors, $\overline{K}$ and $\tilde{K}$, respectively associated to $\overline{\nabla}$ and $\tilde{\nabla}$, we can write:
\begin{description}
\item[$\overline{g}$-Gauss equation]{
\begin{equation}
j^\ast \overline{K} = \tilde{K} + \overline{K}_E
\label{e2.45}
\end{equation}
where: $\overline{K}_E(y,t,v,w) := \phi(v,y)\,\phi(w,t) - \phi(v,t)\,\phi(w,y)$, for $t,y,v,w \in T{\cal V}_3$, plays a role similar to the extrinsic curvature, and the }
\item[$\overline{g}$-Codazzi-Mainardi equation]{
\begin{equation}
\overline{K}(n,t,v,w) = \tilde{\nabla}_v \phi(w,t)-\tilde{\nabla}_w \phi(v,t)
\label{e2.46}
\end{equation}
}
\end{description}


\section{Rigid spatial metrics obtained by constriction \label{S3}}
${\cal E}_3$ is a timelike 3-congruence on which we want to set up a rigid frame of reference, that is, to assign a time scale and a spatial metric.
 
From now on we shall restrict ourselves to the Einsteinian proper time scale:
$\theta:=g(u,\_ )$. It is determined by the spacetime metric and the 3-congruence, and no more ingredients are necessary. The components in a given basis are $\theta_\alpha = g_{\alpha\beta } u^\beta =: u_\alpha$. From the physical viewpoint, given two events $x$, $y$ that happen at the same point $\pi x = \pi y\in {\cal E}_3$, then $\int_{x}^{y} \,\theta$ is the time elapsed between $x$ and $y$ as measured by a stationary standard clock at this point in space.

As for the spatial metric, a rigid $\overline{g}$ can always be associated to a 3-congruence ${\cal E}_3$, at least locally. Indeed, it suffices to assign values for $\overline{g}$ on a hypersurface that is nowhere tangent to the worldlines in ${\cal E}_3$ and extend it according to the transport law ${\cal L}(u)\overline{g} =0$. However, a spatial metric like the latter would rarely have any significance from the physical point of view.

The simplest choice for a physically significant spatial metric is \cite{Born1909}:
\begin{equation}
\overline{g} := \hat{g} \qquad {\rm where} \qquad \hat{g}:= g + \theta\otimes\theta \;,
\label{e3.1}
\end{equation}
that is, the radar or Fermat metric \cite{Landau}. But, as commented in section 1, it is well known \cite{Herglotz} that the latter metric is rigid only for a very short class of congruences. 

Our aim is to find a slight variation of formula (\ref{e3.1}) such that the resulting spatial metric $\overline{g}$ is rigid and has constant curvature. The resulting reference frame $({\cal E}_3,\overline{g},\theta)$ will be therefore rigid and the space of reference will have the property of free mobility \cite{Cartan}. We are also interested in that the class of congruences admitting such a spatial metric is ``large enough'', meaning that given one worldline ${\cal C}$, at least one congruence in this class exists such that: (a) it contains ${\cal C}$ and (b) its vorticity has a preassigned value on ${\cal C}$.

\begin{definition}\label{d4}
Two metrics $\overline{g}$ and $\hat g$ are related through a {\em constriction transformation} if there exist a 1-form $\mu\in [u]^\perp \subset \Lambda^1{\cal V}_4$ and a positive function $\varphi\in\Lambda^0{\cal V}_4$ such that:
\begin{equation}
\overline{g} = \varphi\,(\hat g + \epsilon\,\mu\otimes\mu) \;,\qquad\qquad \epsilon=\pm 1
\label{e3.2}
\end{equation}
\end{definition}
Let $m$ be the vector field resulting from ``raising the index" in $\mu$,
$\,\mu = g(m,\_ ) \;,$ which is obviously $g$-orthogonal to $u$.

\begin{definition}  \label{d4a}
The constriction (\ref{e3.2}) is said to be {\em unitary} if 
$\, \overline{g}(m,m) = \hat g(m,m)$\,.
\end{definition}
The latter condition is equivalent  to the constraint:
\begin{equation}
\varphi\,[1+\epsilon\,g(m,m)] = 1  
\label{e3.5}
\end{equation}
on $\varphi$ and $\mu$.

Definition \ref{d4} amounts to say that $\overline{g}$ can be diagonalized in a $\hat g$-orthonormal basis and one of the principal values is double. If besides the constriction is unitary, then the third principal value is 1.

We want to prove that there is a wide enough class of reference frames such that:
\begin{list}
{(\alph{llista1})}{\usecounter{llista1}}
\item The metric $\overline{g}$ is related with the Fermat metric by a {\it constriction}:
      \begin{equation}
       \overline{g} = \varphi\,(g + \theta\otimes\theta + \epsilon\,\mu\otimes\mu)
       \label{e3.5a}
      \end{equation}
      where $\varphi$ and $\|\mu\|^2:=g^{\alpha\beta}\mu_\alpha\mu_\beta= g(m,m)\,$ fulfill a previously fixed arbitrary constraint: 
      \begin{equation}
      \|\mu\|^2 = \Psi(\varphi) \,.
      \label{e3.8a}
      \end{equation}
      The components of the metric (\ref{e3.5a}) are: $\overline{g}_{\alpha\beta } = \varphi\,(g_{\alpha\beta } +         u_\alpha u_\beta + \epsilon\,\mu_\alpha\mu_\beta )\,$.
\item The metric $\overline{g}$ is rigid:
      \begin{equation}
       S = 0
       \label{e3.6}
      \end{equation}
\item and has constant curvature
      \begin{equation}
       {\cal F} := \overline{K} + k \,\overline{K}^{(0)} = 0
       \label{e3.7}
      \end{equation}
      where $k$ is a constant and 
      \begin{equation}
       \overline{K}^{(0)}(t,y,v,w) :=\overline{g}(t,v)\,\overline{g}(y,w) - \overline{g}(t,w)\,\overline{g}(y,v)
       \label{e3.8}
      \end{equation}
\end{list}

Equations (\ref{e3.6}) and (\ref{e3.7}) provide a partial differential system that we shall solve for the unknowns $u$, $\mu$ and $\varphi$, subject to the constraint (\ref{e3.8a}). We must therefore pose a Cauchy problem by means of a convenient Cauchy hypersurface ${\cal V}_3$ and some Cauchy data on it. 

Since we intend to reconstruct the 3-congruence ${\cal E}_3$ out of a part of it ---a sub\-con\-gruen\-ce---, we shall assume that ${\cal V}_3$ contains all worldlines in ${\cal E}_3$ that have a contact with it. Thus, ${\cal V}_3$ is lined with a 2-subcongruence ${\cal E}_2$. Hence, the Cauchy hypersurface and data will consist, at least, in: (i) a hypersurface: $j:{\cal V}_3 \hookrightarrow {\cal V}_4$
and (ii) a 2-subframe $({\cal E}_2,\tilde{g},\tilde{\theta})$ on ${\cal V}_3$: $\tilde u$ is the $g$-unitary tangent vector, $\tilde{g}$ is a spatial metric (${\rm Rad\,}\tilde{g} = {\rm span}\,[\tilde{u}]$) and $\tilde{\theta} = g(\tilde u,\_ )$ is the Einsteinian proper time.

We seek a reference frame $({\cal E}_3,\overline{g},\theta)$ on ${\cal U}\subset{\cal V}_4$, 
a spacetime neigbourhood of ${\cal V}_3$, such that:
\begin{list}
{(\alph{llista1})}{\usecounter{llista1}}
\item It has $({\cal E}_2,\tilde{g},\tilde{\theta})$ as a subframe, that is:
       $$\forall p\in{\cal V}_3\,,\qquad \tilde u_p = u_p \;,\qquad \tilde{g}_p=j^\ast\overline{g}_p \qquad \qquad \qquad \mbox{ and}$$
\item the spatial metric $\overline{g}$ meets conditions (\ref{e3.8a}), (\ref{e3.6}) and (\ref{e3.7}).
\end{list}

We shall respectively denote by $\zeta$ and $n$ the 1-form orthogonal to ${\cal V}_3$ and the associated  vector field, as introduced in section \ref{SS2.5}\,. We shall also use the same symbols, $n$ and $\zeta$, to denote their extensions to a neighbourhood of ${\cal V}_3$. Besides, we assume that $\langle\zeta,n\rangle = \zeta_\alpha n^\alpha =1$
also hold for the extensions.

The following expressions will be useful later on:
\begin{equation}
\left. \begin{array}{lll}
\displaystyle{\overline{g}_{\alpha\beta } = \varphi\left( \hat g_{\alpha\beta } + \epsilon\,\mu_\alpha\mu_\beta \right)} & \qquad \hat g_{\alpha\beta }:=g_{\alpha\beta } + u_\alpha u_\beta & \qquad \displaystyle{\overline{h}^{\alpha\beta } = \varphi^{-1}\hat g^{\alpha\beta } -\frac{\epsilon\,\mu^\alpha\mu^\beta }{a_1}} \\
\displaystyle{\zeta_\alpha = \varphi\,\left[n_\alpha + \mu_\beta n^\beta \,\mu_\alpha\right]} & \qquad \displaystyle{n^\alpha = \varphi^{-1}\zeta^\alpha -\frac{\epsilon (\mu^\beta \zeta_\beta )}{a_1}\,\mu^\alpha} &  \qquad \zeta_\beta \mu^\beta = \mu_\beta n^\beta \,a_1
\end{array}  \right\}
\label{e3.9}
\end{equation}
where $a_1:=\overline{g}(m,m)=(1+\epsilon\|\mu\|^2)\varphi$. Notice that $\varphi$ and $a_1$ must be positive in order that $\overline{g}$ be a spatial metric. (Recall that indices are raised and lowered with the spacetime metric.)

On its turn, 
\begin{equation}
\Pi^\mu_\nu := \delta^\mu_\nu - n^\mu \zeta_\nu 
\label{e3.10a}
\end{equation}
will denote the projector onto the tangent space $T{\cal V}_3$.

\subsection{Differential system and subsidiary conditions \label{SS3.1}}
In order to see whether ${\cal V}_3$ is a non-characteristic hypersurface for the partial differential system, we must analyse the occurrence of second order normal derivatives  in equations (\ref{e3.6}) and (\ref{e3.7}). For this purpose, it will be most useful to study the components of these equations in a given basis. The right hand side of (\ref{e3.6}) yields:
\begin{equation}
S_{\mu\nu} \equiv \nabla_u \overline{g}_{\mu\nu} + \overline{g}_{\mu\rho} \nabla_\nu u^\rho + \overline{g}_{\rho\nu} \nabla_\mu u^\rho
\label{e3.11}
\end{equation}
Therefore, only the normal contraction:
\begin{equation}
S_{\mu\nu} n^\nu \equiv \nabla_u \overline{g}_{\mu\nu} n^\nu + \overline{g}_{\mu\rho} \nabla_n u^\rho + n^\nu \overline{g}_{\rho\nu} \nabla_\mu u^\rho 
\label{e3.12}
\end{equation}
does depend on normal derivatives of the unknowns (at most of first order), whereas the projection:
\begin{equation}
\Pi^\mu_\alpha \Pi^\nu_\beta S_{\mu\nu} n^\nu = 0
\label{e3.12a}
\end{equation}
only depends on derivatives along directions that are orthogonal to $\zeta$.

To proceed with a similar analysis for (\ref{e3.7}), we first take into account the expression (\ref{e2.208}), where $\overline{K}_{\mu\nu\alpha\beta }$ is split in several parts on the basis of criteria of symmetries and orthogonality relatively to $u^\rho$. It then results that, using also (\ref{e3.6}), equation (\ref{e3.7}) is equivalent to:
\begin{equation}
\left.\begin{array}{rcl}
\Upsilon_{\mu\nu\alpha\beta } &:= & T_{\mu\nu\alpha\beta } + k \overline{K}^{(0)}_{\mu\nu\alpha\beta } = 0 \\
L_{\beta \mu\nu} &:=& 2 u^\alpha \,\overline{K}_{\mu\nu\alpha\beta } = 0
\label{e3.13a}
\end{array}\right\}
\end{equation}
(because $Q_{\mu\nu\alpha\beta }$ already vanishes as a consequence of $S_{\mu\nu} =0$.)

It is thus obvious that (\ref{e3.6}) and (\ref{e3.7}) together are equivalent to
\begin{equation}
\Upsilon_{\mu\nu\alpha\beta } = 0 \;,\qquad \qquad L_{\beta \mu\nu} =0 \;,\qquad \qquad S_{\mu\nu} =0
\label{e3.14}
\end{equation}

Only the first and second equations (\ref{e3.14}) contain second order derivatives of the unknowns. Furthermore, by considering the symmetries of $\Upsilon_{\mu\nu\alpha\beta }$ and $L_{\beta \mu\nu}$\, , we easily conclude that only the contractions:
\begin{equation}
\Upsilon_{\mu\nu\alpha\beta } n^\mu n^\alpha = 0 \qquad {\rm and} \qquad L_{\beta \mu\nu} n^\nu n^\beta =0 
\label{e3.15}
\end{equation}
give rise to a set of independent equations containing second order normal derivatives.

From the symmetries of $\Upsilon_{\mu\nu\alpha\beta }$ and $L_{\beta \mu\nu}$ and their orthogonality to $u^\rho$, it results that each expression in (\ref{e3.15}) yields three independent equations. We have thus six equations for the six independent unknowns $\nabla_n^2 u^\rho$, $\nabla_n^2 \mu_\alpha$ and $\nabla_n^2 \varphi$. (Recall that the unknowns are submitted to the constraints: $g_{\alpha\beta } u^\alpha u^\beta = -1$,  $\mu_\alpha u^\alpha = 0$ and $\|\mu\|^2 = \Psi(\varphi)$.)

In Appendix A it is shown that ${\cal V}_3$ is non-characteristic provided that the Cauchy data are chosen so that
\begin{equation}
 M:=\mu^\alpha \zeta_\alpha \neq 0 \qquad \qquad  a_2:= a_1\|\mu\|^2-(\mu^\alpha \zeta_\alpha)^2 \neq 0 
\label{e3.24}
\end{equation}

\subsection{The subsidiary conditions \label{SS3.2}}
So far we have used part of the components of equations (\ref{e3.14}) to set up the partial differential system (\ref{e3.15}) which has ${\cal V}_3$ as a Cauchy hypersurface. We shall now see what is the role played by the remaining equations (\ref{e3.14}).

Considering the symmetries of the several components of $\Upsilon_{\mu\nu\alpha\beta }$ and $L_{\beta \mu\nu}$ and using that we are dealing with a solution of (\ref{e3.15}), we can write the left hand sides of the first and second equations (\ref{e3.14}) as:
\begin{equation}
\left.\begin{array}{l}
\Upsilon_{\mu\nu\alpha\beta } = t_{\mu\nu\alpha\beta } + 2 \zeta_{[\mu} t_{\nu]\alpha\beta } + 2 \zeta_{[\alpha} t_{\beta ]\mu\nu} \\
L_{\beta \mu\nu} = l_{\beta \mu\nu} + 2 \zeta_{(\beta } l_{\nu)\mu} +\zeta_\mu l^\prime_{\beta \nu}+  2 \zeta_\mu\zeta_{(\beta } l_{\nu)}
\end{array}  \right\}
\label{e3.25}
\end{equation}
where
\begin{equation}
\left.\begin{array}{lll}
t_{\mu\nu\alpha\beta } := \Upsilon_{\lambda\rho\sigma\gamma}\, \Pi^\lambda_\mu\, \Pi^\rho_\nu\, \Pi^\sigma_\alpha\, \Pi^\gamma_\beta  \quad &  t_{\nu\alpha\beta } := \Upsilon_{\mu\nu\rho\sigma}\, n^\mu\,\Pi^\rho_\alpha\,\Pi^\sigma_\beta  \quad &
l_{\nu} := L_{\lambda\sigma\rho}\, n^\sigma n^\rho \Pi^\lambda_\nu \\

l_{\beta \mu\nu}:= L_{\lambda\rho\sigma}\, \Pi^\lambda_\beta \, \Pi^\rho_\mu\, \Pi^\sigma_\nu  &
l_{\nu\mu} := L_{\lambda\rho\sigma}\, n^\sigma \Pi^\rho_\mu \Pi^\lambda_\nu &  l^\prime_{\nu\mu} := L_{\lambda\sigma\rho}\, n^\sigma \Pi^\rho_\mu \Pi^\lambda_\nu 
\end{array}  \right\}
\label{e3.26}
\end{equation}
with  $\Pi^\lambda_\mu := \delta^\lambda_\mu - n^\lambda \zeta_\mu$\,.
 
If conditions (\ref{e3.14}) are to be fulfilled, then each one of the following quantities must vanish separately: 
\begin{equation}
t_{\mu\nu\alpha\beta } = 0 \,, \quad\; t_{\nu\alpha\beta }=0 \,,\quad\;   
l_{\beta \mu\nu} =0  \,,\quad\; l_{\nu\mu} =0 \,,\quad\; l_{\nu} =0 \,,\quad\; l^\prime_{\nu\mu} =0 \quad\; {\rm and} \quad\; S_{\mu\nu} = 0
\label{e3.27}
\end{equation}
These relations do not contain second order normal derivatives and will be taken as subsidiary conditions to be fulfilled on the hypersurface ${\cal V}_3$.

\begin{proposition}  \label{p39}
Let $u^\rho$, $\mu_\alpha$, $\varphi$ be a solution of the partial differential system (\ref{e3.15}) for a set of Cauchy data fulfilling the subsidiary conditions (\ref{e3.27}) on ${\cal V}_3$. Then, these conditions are also met in the neighbourhood of ${\cal V}_3$ where the solution is valid.
\end{proposition}

\paragraph{Proof: }
The second Bianchi identity (\ref{e2.207}) holds whatever is the spatial metric and the associated connexion $\overline{\nabla}$. In particular, it does hold for the metric $\overline{g}_{\mu\nu}$ obtained by introducing the given solution in expression (\ref{e3.5a}). Moreover, from (\ref{e2.20}) and (\ref{e3.8}) it follows that:
$$ \overline{\nabla}_\lambda \overline{K}^{(0)}_{\mu\nu\alpha\beta } = - 2 u_\lambda (S_{\mu[\alpha} \overline{g}_{\beta ]\nu} - \overline{g}_{\mu[\alpha} S_{\beta ]\nu} ) $$
which, combined with (\ref{e2.19}), (\ref{e2.207}) and (\ref{e3.7}), allows to write:
\begin{equation}
\nabla_\lambda {\cal F}_{\mu\nu\alpha\beta } + \nabla_\alpha {\cal F}_{\mu\nu\beta \lambda} + \nabla_\beta {\cal F}_{\mu\nu\lambda\alpha} + {\rm linear}\, ({\cal F}, S) = 0
\label{e3.30}
\end{equation}
where ``${\rm linear}\,({\cal F}, S)$'' stands for terms that depend linearly on ${\cal F}_{\mu\nu\alpha\beta}$ and $S_{\mu\nu}$. 

Contracted with $n^\lambda \Pi^\alpha_\rho \Pi^\beta _\sigma $, the latter equation becomes:
\begin{equation}
\nabla_n[{\cal F}_{\mu\nu\alpha\beta }\Pi^\alpha_\rho \Pi^\beta _\sigma ] + \mbox{l.n.p.t.} = 0
\label{e3.30a}
\end{equation}
where ``l.n.p.t.'' means ``linear non-principal terms'', that is, terms that depend linearly on $t_{\mu\nu\alpha\beta }$\,, $t_{\nu\alpha\beta }\,$, $l_{\beta \mu\nu}\,$, $l_{\nu\mu}\,$, $l^\prime_{\nu\mu}$\,, $l_\nu$ and $S_{\nu\mu}$ and their first order derivatives along directions that are orthogonal to $n^\rho$.

Taking (\ref{e2.208}), (\ref{e3.7}) and (\ref{e3.13a}) into account, we can then write:
\begin{equation}
{\cal F}_{\mu\nu\alpha\beta } = \Upsilon_{\mu\nu\alpha\beta } +Q_{\mu\nu\alpha\beta } -u_{[\alpha} L_{\beta ]\mu\nu}
\label{e3.30b}
\end{equation}
that, substituted into (\ref{e3.30a}) and using (\ref{e3.25}) leads to:
\begin{eqnarray*}
\nabla_n t_{\mu\nu\rho\sigma} + 2\zeta_{[\mu} \nabla_n t_{\nu]\rho\sigma} - 
u_\alpha \Pi^\alpha_{[\rho} \nabla_n l_{\sigma]\mu\nu} - u_\alpha \Pi^\alpha_{[\rho} \nabla_n l_{\sigma]\mu} \zeta_\nu  & & \\  -u_\alpha \Pi^\alpha_{[\rho} \nabla_n l_{\sigma]}\zeta_\mu \zeta_\nu - u_{\alpha} \Pi^\alpha_{[\rho} \nabla_n l^\prime_{\sigma]\nu} \zeta_\mu  + 
\Pi^\alpha_\rho \Pi^\beta _\sigma \nabla_n Q_{\mu\nu\alpha\beta } + \mbox{l.n.p.t.} &=& 0 
\end{eqnarray*}
Now, $\nabla_n Q_{\mu\nu\alpha\beta }$ only contributes terms like $\nabla_n S_{\mu\nu} \Theta_{\alpha\beta }$, where $\Theta_{\alpha\beta }$ is known as far as the solution $u^\alpha$ is given. On its turn, taking (\ref{e2.208}) into account, we easily obtain that:
\begin{equation}
\nabla_n S_{\mu\nu} = n^\lambda H_{\lambda\mu\nu} + \mbox{l.n.p.t.} = n^\lambda L_{\lambda(\mu\nu)} + \mbox{l.n.p.t.} = l_{(\nu\mu)} + \zeta_{(\mu} l^\prime_{\nu)} n^\rho+ \mbox{l.n.p.t.}
\label{e3.31b}
\end{equation}
whence it folowws that $\nabla_n Q_{\mu\nu\alpha\beta } = \mbox{l.n.p.t.}$ and therefore (\ref{e3.30b}) can be written as:
\begin{eqnarray}
& &  \nabla_n t_{\mu\nu\rho\sigma} + 2\zeta_{[\mu} \nabla_n t_{\nu]\rho\sigma} - 
u_\alpha \Pi^\alpha_{[\rho} \nabla_n l_{\sigma]\mu\nu} \nonumber \\ 
& & - u_\alpha \Pi^\alpha_{[\rho} \nabla_n l_{\sigma]}\zeta_\mu \zeta_\nu - u_\alpha \Pi^\alpha_{[\rho} \nabla_n l_{\sigma]\mu} \zeta_\nu - u_{\alpha} \Pi^\alpha_{[\rho} \nabla_n l^\prime_{\sigma]\nu} \zeta_\mu + \mbox{l.n.p.t.} =0 
\label{e3.31a}
\end{eqnarray}
On its turn, from (\ref{e3.31b}) we have that:
\begin{equation}
\nabla_n S_{\mu\nu} + \mbox{l.n.p.t.} = 0
\label{e3.31c}
\end{equation}

Thus, (\ref{e3.31a}) and (\ref{e3.31c}) is a linear, first order, partial differential system on the unknowns $t_{\rho\sigma\alpha\beta }$, $t_{\beta \rho\sigma}$, $l_{\beta \rho\sigma}$,  $l_{\rho\sigma}$, $l^\prime_{\sigma\nu}$, $l_\nu$  and $S_{\mu\nu}$. In Appendix B we prove that the hypersurface ${\cal V}_3$ is non-characteristic. The uniqueness of the solution\footnote{So that,the validity of our results is restricted to the analytical case \cite{John}} together with the vanishing of the variables on ${\cal V}_3$, i. e., conditions (\ref{e3.27}), imply that the latter extend to the neigbourhood where $u^\rho$, $\mu_\alpha$, $\varphi$ are defined. \hfill $\Box$

\subsection{The Cauchy data \label{SS3.5}}
We must now see whether Cauchy data on ${\cal V}_3$ can be found such that the subsidiary conditions (\ref{e3.27}) are met on this hypersurface. From now on and until the end of the present subsection all equalitites must be understood restricted to ${\cal V}_3$.

\begin{proposition} \label{p11}
The following two sets of conditions on ${\cal V}_3$ are equivalent:
\begin{list}
{(\alph{llista})}{\usecounter{llista}}
\item \hspace*{1em} $t_{\mu\nu\alpha\beta }= 0 \,,\qquad t_{\nu\alpha\beta }= 0 \,,\qquad l_{\beta \mu\nu}= 0 \,,\qquad l^\prime_{\mu\nu}= 0 \,,\qquad S_{\mu\nu}= 0$\,, and 
\item \hspace*{1em}for any $t,z,v,w \in T{\cal V}_3$\,,
\begin{eqnarray}
{\cal F}(t,z,v,w) = 0 \,, &\qquad &  \qquad {\cal F}(n,z,v,w) = 0 
\label{e3.104} \\
S(v,w) = 0 \,, &\qquad &  \qquad S(n,v) = 0 
\label{e3.105}
\end{eqnarray}
\end{list}
\end{proposition}

\paragraph{Proof: }
From (\ref{e3.30b}) and (\ref{e3.25}) we have that $\forall t,z,v,w \in{\cal V}_3$\,:
\begin{eqnarray*}
{\cal F}(t,z,v,w) &= &(t_{\mu\nu\alpha\beta } + Q_{\mu\nu\alpha\beta }-u_{[\alpha} l_{\beta ]\mu\nu})t^\mu z^\nu v^\alpha w^\beta \qquad {\rm and} \\
{\cal F}(n,z,v,w) &= &(t_{\nu\alpha\beta } + n^\mu Q_{\mu\nu\alpha\beta }- u_{[\alpha}l^\prime_{\beta ]\nu} ) z^\nu v^\alpha w^\beta   
\end{eqnarray*}
Whence the proof follows straightforward after taking into account the expression (\ref{e2.209}) for $Q_{\mu\nu\alpha\beta }$\,. \hfill $\Box$

\begin{proposition} \label{p11a}
If $S_{\alpha\beta }=0$ and $l^\prime_{\alpha\beta }=0$ on ${\cal V}_3$, then $l_{\nu\mu} = l_\mu =0$ on ${\cal V}_3$.
\end{proposition}

Indeed, from (\ref{e3.26}) and (\ref{e2.209}) we have that:
\begin{eqnarray*}
l_{\nu\mu} &=& - l^\prime_{\nu\mu} + 2 H_{\lambda\rho\sigma} n^\sigma \Pi^\rho_\mu \Pi^\lambda_\nu   \\
l_\nu &=& L_{\lambda\sigma\rho} \Pi^\lambda_\nu n^\sigma n^\rho = H_{\lambda\sigma\rho} n^\sigma n^\rho \Pi^\lambda_\nu 
\end{eqnarray*}
that vanishes because the right hand sides only contain tangential derivatives of $S_{\alpha\beta }$. \hfill $\Box$

The first equation (\ref{e3.104}) amounts to $j^\ast{\cal F} =0$. Therefore, using (\ref{e2.45}), (\ref{e2.46}) and (\ref{e3.7}) we have that:
\begin{eqnarray}
 & & \tilde{K} + K_E + k \tilde{K}^{(0)} = 0  
\label{e3.106} \\
{\rm and} & & 
\tilde{\nabla}_v \phi(w,z)-\tilde{\nabla}_v \phi(w,z)  \,, \qquad \qquad \forall v, w, z\in T{\cal V}_3
\label{e3.107}
\end{eqnarray}

On its turn, equations (\ref{e3.105}) can be written as:
\begin{equation}
\tilde{S} := j^\ast S = {\cal L}(\tilde{u})\tilde{g} = 0  \qquad {\rm and} \qquad 
S(n,\_) = 0
\label{e3.108}
\end{equation}

To prove that we can indeed choose a set of Cauchy data on ${\cal V}_3$ fulfilling simultaneously the conditions (\ref{e3.106}-\ref{e3.108}), it is enough to realize that, if we assume $\phi = 0$\,, then:
\begin{list}
{(\alph{llista})}{\usecounter{llista}}
\item the condition (\ref{e3.107}) is authomatically warranted,
\item $K_E = 0$, and conditions (\ref{e3.106}) and (\ref{e3.108}) lead to:
      \begin{equation}
      \tilde{K} + k \tilde{K}^{(0)} = 0 \,, \qquad \tilde{S} = 0 \,,
      \label{e3.110}
      \end{equation}
      which is precisely the same problem we are studying but on a manifold ${\cal V}_3$ with a lower number of dimensions, and
\item the three remaining conditions:
      \begin{equation}
      \phi = 0 \,,  \quad {\rm and} \qquad  S(n,\_) = 0 
      \label{e3.111}
      \end{equation}
together with the algebraic constraint $\|\mu\|^2=\Psi(\varphi)$, allow to determine the normal derivatives $\nabla_n \mu_\alpha$\,, $\nabla_n u^\rho$ and  $\nabla_n\varphi$ in terms of the values of $\mu_\alpha$\,, $u^\rho$ and $\varphi$ on ${\cal V}_3$, as we prove in detail in appendix C.
\end{list}

The discussion developed so far can be summarized in the following 

\begin{theorem} \label{t10}
Let $({\cal V}_4,g)$  and $j:{\cal V}_3 \hookrightarrow{\cal V}_4$ be respectively a spacetime with constant signature $(+3,-1)$ and a submanifold such that $j^\ast g$ has constant signature $(+2,-1)$. Let $\Psi(\varphi)$ be a given positive function. Besides, let $({\cal E}_2,\tilde{g},\tilde\theta)$ be an Einsteinian reference frame on $({\cal V}_3,j^\ast g)$ such that:
\begin{list}
{(\alph{llista})}{\usecounter{llista}} 
\item the spatial metric is rigid and has constant curvature: $\tilde{K} + k \tilde{K}^{(0)}=0$, and
\item $\tilde{g}$ is a constriction of $j^\ast\hat{g}:=j^\ast g + \tilde\theta\otimes\tilde\theta$, that is, there exist $\tilde\varphi \in \Lambda^0{\cal V}_3$ and $\tilde\mu \in \Lambda^1{\cal V}_3$, with $\tilde\mu_p\neq 0$  and 
 $\|\tilde\mu_p\|^2<\Psi(\tilde\varphi_p)$, $\forall p\in{\cal V}_3$, such that: $\tilde{g} = \tilde\varphi[j^\ast g + \tilde\theta\otimes\tilde\theta +\epsilon \tilde\mu\otimes\tilde\mu]$.
\end{list}
Then, there exists an Einsteinian reference frame $({\cal E}_3,\overline{g},\theta)$ on a neighbourhood ${\cal U}$ of ${\cal V}_3\subset {\cal V}_4$, such that:
\begin{list}
{(\alph{llista})}{\usecounter{llista}}
\item  $({\cal E}_2,\tilde{g},\tilde\theta)$ is a subframe of $({\cal E}_3,\overline{g},\theta)$ with second fundamental form $\phi=0$,
\item $\overline{g}$  is rigid and has constant curvature: $\overline{K} + {k} \overline{K}^{(0)}=0$, and
\item $\overline{g}$ is a constriction of $\hat{g}:= g + \theta\otimes\theta$, that is, there exist $\varphi \in \Lambda^0{\cal U}$ and $\mu \in \Lambda^1{\cal U}$ such that: $\overline{g} = \varphi[g + \theta\otimes\theta +\epsilon \mu\otimes\mu]$ and $\|\mu\|^2=\Psi(\varphi)$, with $\tilde\varphi = j^\ast\varphi$ and $\tilde\mu = j^\ast\mu$.
\end{list}
\end{theorem}

\paragraph{Proof: }
According to the hypothesis, $({\cal E}_2,\tilde{g},\tilde\theta)$ satisfies condition (\ref{e3.110}). We then choose 
$$\tilde{M}=\sqrt{\frac{a_1}{2}\left(\sqrt{1+\frac{4[\Psi(\tilde\varphi)- \|\tilde\mu\|^2]}{a_1\tilde\varphi}}-1\right)} 
$$
and take
$$ \varphi=\tilde\varphi \,, \qquad \mu = \tilde\mu + \tilde{M}\zeta \,, \qquad u = \tilde u \qquad {\rm on} \quad {\cal V}_3 \,,$$
We then set $\phi =0$, that authomatically satisfies (\ref{e3.107}), and use (\ref{e3.111}) to determine $\nabla_n u^\rho$, $\nabla_n \mu_\alpha$ and $\nabla_n \varphi$ on ${\cal V}_3$ according with the discussion in subsection \ref{SS3.5} and  Appendix C.

With the Cauchy data so chosen, ${\cal V}_3$ is a non-characteristic hypersurface for the partial differential system (\ref{e3.15}) according to the discussion in subsection \ref{SS3.1} and in Appendix A. Let $u^\rho$, $\varphi$ and $\mu_\alpha$ be a solution. Since the Cauchy data have been chosen so that the subsidiary conditions (\ref{e3.27}) are fulfilled on ${\cal V}_3$, according to the discussion in subsection \ref{SS3.2} and Appendix B, they will be also met in a neighbourhood of ${\cal V}_3$.

Whence it follows that the congruence defined by the vector field $u$ and the metric $\overline{g} = \varphi[g + \theta\otimes\theta +\epsilon \mu\otimes\mu]$ and the Einsteinian time scale $\theta$ define the sought reference frame. \hfill $\Box$

\subsection{The problem in 2+1 dimensions \label{SS3.6}}
Theorem \ref{t10} reduces the proof of the existence of the reference frame $({\cal E}_3,\overline{g},\theta)$ mentioned at the begining of the present section to proving the existence of the subframe in the hypothesis. We shall now prove that the latter does indeed exist.

\begin{proposition} \label{ta}
Let $({\cal V}_3, g)$ and ${\cal E}_1$ be respectively a (2+1)-spacetime and a 1-congruence of timelike worldlines. Furthermore, let ${\cal V}_2$ be the submanifold spanned by ${\cal E}_1$ and $\tilde{u}$ be the unitary tangent vector field. Then, there exists a unitary timelike vector field $u$ in a neigbourhood ${\cal U}_3\subset{\cal V}_3$ of ${\cal V}_2$ such that the Fermat tensor is conformally rigid, i. e.,
\begin{equation}
\exists \gamma \qquad \mbox{such that} \qquad {\cal L}(u)\hat{g} =\gamma \hat{g}
\label{A2}
\end{equation}
and for all $p\in{\cal V}_2$: $u_p = \tilde{u}_p$, that is ${\cal E}_1$ is a subcongruence of ${\cal E}_2$.
\end{proposition}

\paragraph{Proof: }
Written in any basis, equation (\ref{A2}) reads:
$$\hat{g}^\lambda_\alpha \hat{g}^\mu_\beta (\nabla_\lambda u_\mu + \nabla_\mu u_\lambda) = \gamma \hat{g}_{\alpha\beta } $$
which is equivalent to the vanishing of the traceless part of the right hand side:
\begin{equation}
(\hat{g}^\lambda_\alpha \hat{g}^\mu_\beta + \hat{g}^\mu_\alpha \hat{g}^\lambda_\beta - \hat{g}^{\lambda\mu} \hat{g}_{\alpha\beta } )\nabla_\lambda u_\mu = 0
\label{A4}
\end{equation}

Now, let $n$ be a unitary vector field on ${\cal V}_3$ such that $n_p$ is orthogonal to $T_p{\cal V}_2$ for all $p\in{\cal V}_2$, and let $\{ e_0,e_1,e_2\}$ be an orthonormal basis on a neigbourhood of ${\cal V}_2$ such that $e_2=n$ and $e_0 = \tilde{u}$ on ${\cal V}_2$. In this basis we have that $\hat{g}^\lambda_\alpha = \delta^\lambda_\alpha - \delta^0_\alpha \delta^\lambda_0 $ on ${\cal V}_2$, and equations (\ref{A4}) yield:
\begin{equation}
\nabla_1 u_2 = -\nabla_2 u_1 \qquad {\rm and} \qquad  \nabla_1 u_1 = \nabla_2 u_2 \qquad \mbox{on }\;{\cal V}_2 \,.
\label{A5}
\end{equation}
Whence it follows that ${\cal V}_2$ is non-characteristic for the partial differential system (\ref{A4}). The Cauchy Kobalewski theorem \cite{John} can then be invoked to end the proof. \hfill $\Box$

As a consequence, a conformal factor $\varphi$ can be found such that $\tilde{g}^\prime:= \tilde\varphi\,\hat g$ is rigid for the congruence ${\cal E}_2$. Indeed, from (\ref{A2}) and the rigidity of $\tilde{g}^\prime$ it immediately follows that: 
$(u\tilde\varphi) \,\hat{g} + \tilde\varphi\,{\cal L}(u)\hat g = 0$, 
whose trace yields a differential equation on $\tilde\varphi$
\begin{equation}
u (\log\tilde\varphi) = -\frac12 g^{\alpha\beta }{\cal L}(u)\hat{g}_{\alpha\beta }
\label{A4a}
\end{equation}

Now consider the spatial metric $\tilde{g}^\prime :=\tilde\varphi\hat{g}$. By Theorem \ref{t1} $\tilde{g}^\prime$ can be projected onto a Riemannian metric\footnote{Since all the results proved in section \ref{S2} do not depend on the number of dimensions, they also apply in the present case} $g_2^\prime$ on ${\cal E}_2$ such that $\pi^\ast g_2^\prime = \tilde{g}^\prime$. ($\pi: {\cal V}_3\longrightarrow {\cal E}_2$ denotes the canonical projection.) Now, $g_2^\prime$ is a Riemannian metric on the 2-manifold ${\cal E}_2$ and, applying Lemma \ref{l20} in Appendix D, a non-vanishing 1-form $\,\stackrel{2}{\mu}\in\Lambda^1{\cal E}_2$ can be chosen so that the Riemannian metric $g_2:= g_2^\prime + \epsilon\stackrel{2}{\mu}\otimes\stackrel{2}{\mu}$ has constant curvature.

Finally, the spatial metric $\tilde{g} := \pi^\ast g_2$ on ${\cal V}_3$ is rigid (by construction) and has constant curvature. Moreover, it is a constriction of the Fermat metric $\hat{g}$ because:
$$\tilde{g} = \pi^\ast\left(g^\prime_2 + \epsilon\stackrel{2}{\mu}\otimes\stackrel{2}{\mu}\right) =
\tilde\varphi\left(\hat{g} + \epsilon\tilde\mu\otimes\tilde\mu\right) \qquad {\rm with} \qquad 
\tilde\mu :=\tilde\varphi^{-1/2} \pi^\ast \stackrel{2}{\mu} $$

\section{Conclusion and outlook  \label{S4}}
We shall finish by reviewing the amount of arbitrariness that is left in the derivation of the congruence ${\cal E}_3$ and the spatial metric $\overline{g}$. It will provide an idea of how large is the class of reference frames selected by the condition that the Fermat metric can be transformed by constriction into a rigid, free mobile metric. We shall then examine whether that arbitrariness is enough to allow one of the desiderata invoked in the introduction, namely, that there is at least one of these congruences containing a given timelike worldline and having a prescribed vorticity on it. 

The derivation of the timelike congruence ${\cal E}_3$ and the spatial metric $\overline{g}$ in section \ref{S3} is based in a sequential application of Proposition \ref{ta}, Lemma \ref{l20} and Theorem \ref{t10}. Their hypothesis imply the following list of arbitrary choices:
\begin{list}
{(\alph{llista})}{\usecounter{llista}}
\item A sequence of timelike submanifolds: ${\cal V}_2 \stackrel{j^\prime}{\hookrightarrow} {\cal V}_3 \stackrel{j}{\hookrightarrow} {\cal V}_4 $
where subindices indicate the respective numbers of dimensions.
\item A one-parameter congruence of timelike worldlines, ${\cal E}_1$, on ${\cal V}_2$.
\item A 1-form $\stackrel{2}\mu \in\Lambda^1{\cal E}_1$.
\item The secon fundamental form $\phi$ for $({\cal E}_2,\tilde{g},\tilde\theta)$ as a subframe of $({\cal E}_3,\overline{g},\theta)$. (Although in Theroem \ref{t10} it has been chosen $\phi=0$, other choices are also consistent with equation (\ref{e3.107}), e. g., $\phi = k_1 \tilde{g}$, with $k_1$ constant.)
\item The positive function $\Psi(\varphi)$.
\end{list}
It is obvious to conclude that the class of reference frames obtained contains the class of Born-rigid motions and is larger by far. Indeed, for a Born-rigid motion, ${\cal L}\hat{g} =0$, thus $\hat{g}$ is projectable and a Riemannian metric $\stackrel{3}{g}$ on the space of reference ${\cal E}_3$ exists such that $\pi^\ast\stackrel{3}{g}=\hat{g}$. Now, as an apllication of the results in ref. \cite{CoLlS}, $\stackrel{3}{g}$ can be transformed by constriction into a constant curvature metric on the space. Finally, $\overline{g}$ is obtained by pulling the latter bcak to ${\cal V}_4$. 

Let us now analyse whether the vorticity can be arbitrarily prescribed at one among  the given worldlines (or at the whole submanifold ${\cal V}_2$ spanned by the worldlines in ${\cal E}_1$). First we take an orthonormal tetrad of vectors $\{e_0, e_1, e_2, e_3\}$, that is adapted to the data, namely, the 1-congruence ${\cal E}_1$ and the submanifolds ${\cal V}_2$ and ${\cal V}_3$:
$$ u= e_0 \quad {\rm on}\;\; {\cal V}_2\,, \qquad e_2 \quad {\rm and}\quad e_3 \quad\mbox{are respectively normal to} \quad {\cal V}_2 \;\; {\rm and} \;\; {\cal V}_3 $$
We denote by $\nabla^\prime$ the Riemannian connexion on ${\cal V}_3$. 
The vorticity components are $ \Theta_{ij} = \nabla_i u_j - \nabla_j u_i \,,$ \,$i,j = 1,2,3 $, and denoting by $\Phi$ the extrinsic curvature of ${\cal V}_3$, we can write: 
$$ \Theta_{3a} = \nabla_3 u_a - \Phi_{a0} \,, \qquad a=1,2  
\qquad {\rm on}\;{\cal V}_3           $$
Now, $\nabla_3 u_a$ is given by equation (\ref{e3.114}) and depends on the values of the arbitrary data $\varphi$ and $\mu_1$ on ${\cal V}_2$ [item (c) in the above list], and the extrinsic curvature can also be tuned by choosing a convenient submanifold ${\cal V}_3$. 

As for the remaining non-vanishing vorticity component, on ${\cal V}_3$  we have that: $\Theta_{12} = \nabla_1 u_2 - \nabla_2 u_1 = \nabla^\prime_1 u_2 - \nabla^\prime_2 u_1$, and considering the extrinsic curvature $\Phi^\prime$ of ${\cal V}_2$ as a submanifold of ${\cal V}_3$ we can write: 
$$ \Theta_{12} = - \nabla^\prime_2 u_1 + \Phi^\prime_{10} 
\qquad {\rm on}\;{\cal V}_2         $$
Again, $ \nabla^\prime_2 u_1$ is given by equation (\ref{A5}) as a function of the data on ${\cal V}_2$ and $\Phi^\prime_{10}$ can be conveniently tuned by the choice of the submanifold ${\cal V}_2$. 

It thus seems that enough freedom is left that allows to prescribe the vorticity in the whole ${\cal V}_2$, which is much more than our desideratum invoked at the introduction. Hence, the congruence ${\cal E}_3$ is not determined by the motion of an origin point plus the angular velocity around that origin.

Let us now discuss the physical significance of the notions developed so far. 
When a rigid reference frame $({\cal E}_3,\overline{g},\theta)$ is chosen to describe a specific physical reality, it is actually pressumed that reference bodies and standard rods are available such that they embody this space ${\cal E}_3$ and this metric $\overline{g}$. We are in fact making the hypothesis that ther are some physical bodies that approximately behave in some prescribed way. 

Unfortunately, the nicest choice for the spatial metric $\overline{g}$, namely, $\overline{g} = \hat g$ (the Fermat metric) is not viable, as commented in the introduction. It therefore seems reasonable to change the least and look for $\overline{g}$ keeping a simple prescribed relationship with $\hat g$. We have here considered the case of constriction transformations (definition \ref{d4}), but other possible relationships have also been considered in the literature (\cite{Bel96} and \cite{LloSol} among others). 

Although, on theoretical grounds, prefering one rather than another is a matter of taste, the choice of $\overline{g}$ has measurable consequences. Indeed, given a rigid reference frame, there are two distinguished spatial metrics, first, the one belonging to the reference frame $\overline{g}$ , which is materialized by the reference body and is rigid, and second, the Fermat metric $\hat{g}$, which is embodied in radar signals and is not generally preserved. So that, experiments can be devised to compare both metrics. This is actually the aim of Michelson-Morley type experiments \cite{Brillet}: measuring the anisotropy of the speed of light in vacuum actually compares two standards of length ---one based on light signals and another based on the reference body--- along different space directions. In other words, it amounts to check whether $\hat{g}$ and $\overline{g}$ are or are not conformal to each other. 

\section*{Appendix}
\subsection*{Appendix A: Non-characteristic hypersurfaces for (\ref{e3.15})}
In order to determine under what conditions the hypersurface ${\cal V}_3$ is non-characteristic for the partial differential system (\ref{e3.15}), we have to study its principal part \cite{John}, but instead of checking that the characteristic determinant does not vanish on ${\cal V}_3$, we shall try to solve (\ref{e3.15}) for the second normal derivatives $\nabla_n^2 u^\rho$, $\nabla_n^2 \mu_\alpha$ and $\nabla_n^2 \varphi$ on the Cauchy hypersurface.

Hereafter, and until the end of the present appendix, only the principal part of most  expressions will be explicitly written. Moreover, all equalitites must be understood restricted to ${\cal V}_3$. Thus, using (\ref{e3.12}) and (\ref{e2.202}) and taking the symmetries of $\Upsilon_{\mu\nu\alpha\beta }$ into account, we obtain that the left hand side of the first equation (\ref{e3.15}) is:
\begin{equation}
\Upsilon_{\mu\nu\alpha\beta } n^\mu n^\alpha \cong -\frac12\,\hat\Pi^\lambda_\nu\, \hat\Pi^\sigma_\beta \nabla_n^2 \overline{g}_{\lambda\sigma}
\label{e3.16}
\end{equation}
where $\cong$ stands for ``equal modulo terms that do not depend on second order normal derivatives'' and 
$$\hat{\Pi}^\nu_\mu = \hat{g}^\nu_\rho \Pi^\rho_\mu = \delta^\nu_\mu +u^\nu u_\mu - n^\nu \zeta_\mu \,.$$
Taking then (\ref{e3.5a}) into account, we have that the first equation (\ref{e3.15}) can be written as:
\begin{equation}
\frac12\,\left(\nabla_n^2 \varphi \varphi^{-1} \overline{g}_{\lambda\sigma} + \epsilon \varphi [\mu_\lambda \nabla_n^2 \mu_\sigma + \mu_\sigma \nabla_n^2 \mu_\lambda]  \right)\, \hat\Pi^\lambda_\nu\,\hat\Pi^\sigma_\beta = A_{\nu\beta }
\label{e3.17}
\end{equation}
on ${\cal V}_3$, where $A_{\nu\beta }$ does not depend on second order normal derivatives.

We then proceed similarly with the left hand side of the second expression (\ref{e3.15}). Taking (\ref{e2.209}) and (\ref{e2.19}) into account, and the fact that $n^\rho u_\rho = 0$ on ${\cal V}_3$, we obtain:
$$ L_{\beta \mu\nu} n^\nu n^\beta \cong 2 \nabla_n (S_{\mu\beta } n^\beta ) - \nabla_\mu (S_{\nu\beta } n^\nu n^\beta ) $$
which, using (\ref{e3.12}), leads to: $L_{\beta \mu\nu} n^\nu n^\beta \cong 2 \overline{g}_{\mu\rho} \nabla^2_n u^\rho\,$. Therefore, the second equation (\ref{e3.15}) can be written as:
\begin{equation}
\overline{g}_{\mu\rho} \,\nabla_n^2 u^\rho = A_\mu  \qquad \qquad {\rm on}\quad {\cal V}_3
\label{e3.18}
\end{equation}
where $A_\mu$ does not contain second order normal derivatives.

Finally, the second normal derivative of the algebraic constraint $\|\mu\|^2=\Psi(\varphi)=0$ yields:
\begin{equation}
-\Psi^\prime(\varphi) \nabla_n^2 \varphi + 2  \mu^\beta \nabla_n^2 \mu_\beta = A
\label{e3.19}
\end{equation}
where $A$ does not contain second order normal derivatives and $\Psi^\prime=d\Psi/d\varphi$. 

Let us now proceed to solve equations (\ref{e3.17}), (\ref{e3.18}) and (\ref{e3.19}) for the second normal derivatives of the unknowns. From (\ref{e3.18}) we have that:
\begin{equation}
\nabla_n^2 u^\rho = \overline{h}^{\rho\mu} A_\mu + C u^\rho
\label{e3.20}
\end{equation}
and the coefficient $C$ is determined by the additional condition $u_\rho \nabla_n^2 u^\rho + \nabla_n u_\rho \nabla_n u^\rho =0$, which follows from the constraint $u^\rho u_\rho = -1$\,.

On its turn, by contracting (\ref{e3.17}) respectively with $\overline{h}^{\nu\beta}$ and with $\mu^\nu \mu^\beta $ we obtain that:
\begin{equation}
\left. \begin{array}{l}
\varphi^{-1}\nabla_n^2 \varphi + \epsilon\varphi\,\Pi^\lambda_\alpha \mu^\alpha
\nabla_n^2 \mu_\lambda= A_{\nu\beta } \overline{h}^{\nu\beta } \\
\varphi^{-1}\,a_2\, \nabla_n^2 \varphi + 2\epsilon \varphi\, a_2\, \Pi^\lambda_\alpha \mu^\alpha \nabla_n^2 \mu_\lambda = 2 A_{\nu\beta } \mu^\nu \mu^\beta 
\end{array} \right\}
\label{e3.21}
\end{equation}
with $\, a_2:= a_1\,\|\mu\|^2 - M^2\;$ and $\,M:= \mu^\alpha \zeta_\alpha $.

Then the two equations (\ref{e3.21}) can be solved for 
$$\nabla_n^2 \varphi = D_1 \qquad {\rm and} \qquad (\mu^\lambda -M n^\lambda) \nabla_n^2 \mu_\lambda = D_2\,, $$ 
(where $D_1$ and $D_2$ do not depend on secon order normal derivatives) 
provided that the determinant does not vanish, that is, $a_2 \neq 0$. Combining then the latter equations with (\ref{e3.19}) and provided that $M\neq 0$ we can solve them for $n^\lambda \nabla_n^2 \mu_\lambda$.

After that, we substitute these solutions in the equation that is obtained by contracting (\ref{e3.17}) with $\mu^\beta $ and arrive at: 
$$  a_2 \hat\Pi^\lambda_\nu \nabla^2_n \mu_\lambda = B_\nu  \,,$$
where $B_\nu$ does not contain second order normal derivatives. Now, provided that $a_2 \neq 0$, we can solve it to obtain
$$\nabla_n^2 \mu_\lambda = \frac1{a}\, B_\nu + E_1 u_\nu + E_2 \zeta_\nu  $$
where the coefficients $E_1$ and $E_2$ can be respectively determined from the constraints $\mu_\nu u^\nu=0$ and the value obtained above for $n^\lambda \nabla_n^2 \mu_\lambda$.

To summarize, the hypersurface ${\cal V}_3$ is non-characteristic if 
\begin{equation}
a_2 := a_1\|\mu\|^2 - M^2 \neq 0 \qquad {\rm and}  \qquad M:= \mu^\alpha \zeta_\alpha \neq 0
\label{e3.22}
\end{equation}

\subsection*{Appendix B: Subsidiary conditions}
In order to see whether ${\cal V}_3$ is a non-characteristic hypersurface for the partial differential system (\ref{e3.31a}) and (\ref{e3.31c}), we have to check that it can be solved for the normal derivatives $\nabla_n$ of all the unknowns on ${\cal V}_3$. We shall use that $n^\mu u_\mu = \zeta_\mu u^\mu = 0$, on ${\cal V}_3$, and therefore $u_\alpha \Pi^\alpha_\rho = u_\rho$. From now on all equalities are understood restricted to ${\cal V}_3$. 

 Now, by appropriately projecting successively (\ref{e3.31a}) on the directions of $u^\rho$, $n^\mu$ and also on the transverse directions, we finally obtain after a little algebra that:
$$
\left.\begin{array}{lll} 
\nabla_n t_{\mu\nu\rho\sigma}  + \mbox{l.n.p.t.} = 0  & \qquad
\nabla_n t_{\nu\rho\sigma}   + \mbox{l.n.p.t.} = 0 &  \qquad    \\
\nabla_n l_{\sigma\mu\nu}  + \mbox{l.n.p.t.} = 0 & \qquad
\nabla_n l_{\sigma\mu} + \mbox{l.n.p.t.} = 0 & \qquad         \\
\nabla_n l_{\mu} + \mbox{l.n.p.t.} = 0  & \qquad
\nabla_n l^\prime_{\sigma\nu} + \mbox{l.n.p.t.} = 0 & \qquad
\nabla_n S_{\nu\mu} + \mbox{l.n.p.t.} = 0
\end{array} \right\}
$$
which is already in normal form. Therefore, ${\cal V}_3$ is non-characteristic.

\subsection*{Appendix C: First order normal derivatives }
In order to determine the normal derivatives $\nabla_n \mu_\alpha$\,, $\nabla_n u^\rho$ and  $\nabla_n\varphi$ in terms of the values of $\mu_\alpha$\,, $u^\rho$ and $\varphi$ on ${\cal V}_3$, 
we need to explicitly write the normal derivatives in the three equations (\ref{e3.111}). Thus, taking (\ref{e3.5a}), (\ref{e2.42.3}) and (\ref{e2.42.4}) into account, $\phi_{\mu\nu}= 0$ we have that:
\begin{equation}
\frac12\,\left(\nabla_n \varphi \varphi^{-1} \overline{g}_{\lambda\sigma} + \epsilon \varphi [\mu_\lambda \nabla_n \mu_\sigma + \mu_\sigma \nabla_n \mu_\lambda]  \right)\, \Pi^\lambda_\mu\,\Pi^\sigma_\nu = C_{\mu\nu}
\label{e3.112}
\end{equation}
where $C_{\mu\nu}$ does not depend on normal derivatives. Also, from the algebraic constraint $\|\mu\|^2=\Psi(\varphi)$ we obtain
\begin{equation}
-\Psi^\prime(\varphi)\,\nabla_n\varphi + 2 \mu^\beta \nabla_n \mu_\beta = 0
\label{e3.113}
\end{equation}
The left hand sides of (\ref{e3.112}) and (\ref{e3.113}) look exactly like (\ref{e3.17}) and (\ref{e3.19}), with $\nabla_n$ instead of $\nabla_n^2$. Therefore, similarly as in Appendix A, the linear system (\ref{e3.112}) and (\ref{e3.113}) can be solved for $\nabla_n\mu_\alpha$ and $\nabla_n \varphi$. 

Let us now develop the equation $n^\rho S_{\mu\rho}=0$. From (\ref{e3.12}) we have that:
$$\overline{g}_{\mu\rho} \nabla_n u^\rho + \zeta_\rho \nabla_\mu u^\rho = C_\mu $$
where $C_{\mu}$ does not depend on normal derivatives. A short calculation yields that:
$$\overline{g}_{\mu\rho} \nabla_n u^\rho  = C_\mu - \frac12 C_\alpha n^\alpha \zeta_\mu $$
that, combined with $u_\rho \nabla_n u^\rho =0$, leads to
\begin{equation}
\nabla_n u^\rho  = \overline{h}^{\rho\mu} C_\mu - \frac12 C_\alpha n^\alpha n^\rho 
\label{e3.114}
\end{equation}

\subsection*{Appendix D: Deformation of a Riemannian metric in 2 dimensions}
\begin{lemma}  \label{l20}
Let $({\cal E}_2,g)$ be a Riemannian 2-manifold. A non-vanishing 1-form $\mu\in \Lambda^1{\cal E}_2$ can be locally found such that the deformed metric $g^\prime := g + \epsilon\mu\otimes\mu$ has constant curvature. 
\end{lemma}

\paragraph{Proof: }
Let $\nabla$ and $\nabla^\prime$ be the respective Riemannian connexions. The difference tensor $b_{ij|k} := g^\prime_{kl}[\gamma^{\prime l}_{ij} -\gamma^l_{ij}]$, $\,i,j,k\ldots = 1,2$ is:
\begin{equation}
b_{ij|k} =\frac\epsilon2\,\left[ \nabla_i\left(\mu_j\mu_k\right) + \nabla_j\left(\mu_i\mu_k\right) - \nabla_k\left(\mu_i\mu_j\right)\right] 
\label{ed.1}
\end{equation}
Writing then $b^l_{ij} =\gamma^{\prime l}_{ij} -\gamma^l_{ij}$, the relation between the respective Riemann tensors is
\begin{equation}
R^\prime_{ijkl}=g^\prime_{is} R^s_{\;jkl} + 2 \nabla_{[k}b_{l]j|i} + 2 b^s_{j[k} b_{l]i|s}
\label{ed.2}
\end{equation}
and the condition of constant curvature reads:
\begin{equation}
R^\prime_{ijkl}= k \left(g^\prime_{ik} g^\prime_{jl} - g^\prime_{il} g^\prime_{jk} \right)
\label{ed.3}
\end{equation}
Substituting (\ref{ed.3}) into (\ref{ed.2}) we obtain a tensor differential system on $\mu_i$. Due to the symmetries of Riemann tensors and the fact that we are in 2 dimensions, it results in only one independent component, which is equivalent to the scalar equation obtained on contraction with $g^{ik} g^{jl}$, that is,
\begin{equation}
2 k (1+\epsilon \|\mu\|^2) = R + \epsilon \mu_i \mu_j R^{ij} + \nabla_i\left[g^{jl} g^{ki}(b_{jl|k} - b{jk|l})\right] + g^{jl} g^{ki}\,[b^s_{il} b_{jk|s} - b^s_{jl} b_{ki|s}]
\label{ed.4}
\end{equation}
Since there is only one equation for two unknowns, the problem is underdetermined. We can arbitrarily choose $ \mu_i = f m_i$, where $m_i$ is given and $ m_i m^i = 1 $. With a choice like that, equation (\ref{ed.4}) becomes:
\begin{equation}
\nabla_i \nabla^i f + p^{ij} \nabla_i\nabla_j f + p^i \nabla_i f + p\, f  + q^o f\nabla_i f + q \, f^2 + t = 0
\label{ed.5}
\end{equation}
where $p^{ij}$, $p^i$, $p$, $q^i$, $q$ and $t$ are known functions that depend on the chosen $m_j$ and its derivatives.

This equation being already in normal form, any curve ${\cal E}_1 \subset {\cal E}_2$ that is regular enough is a good Cauchy hypersurface. The Cauchy data are $f $ and $\nabla_n f$ on ${\cal E}_1$.
\hfill $\Box$.

\end{document}